\documentclass{svproc}
\usepackage{url}

\usepackage{amsmath}
\usepackage{amssymb}
\usepackage{graphicx}
\usepackage{microtype}

\usepackage[linesnumbered,ruled,vlined]{algorithm2e}
\begin{document}
\mainmatter              
\title{Profit Maximization in Closed Social Networks}
\titlerunning{Profit Maximization in Closed Social Networks}  
%
\author{Poonam Sharma\inst{} \and Suman Banerjee\inst{} }
\authorrunning{Sharma and Banerjee} 
%
\institute{Department of Computer Sciene and Engineering,\\ Indian Institute of Technology Jammu, Jammu-181221,J\& K, India.\\
\email{\{poonam.sharma,suman.banerjee\}@iitjammu.ac.in}.
}

\maketitle              

\begin{abstract}
Diffusion of information, innovation, ideas, etc. is one of the important phenomena of a social network. Hence, information propagates through the network and reaches from one person to the next. Many times, it is meaningful to restrict the diffusion for a node to a limited number of its neighbors instead of all.  Such social networks are called \emph{Closed Social Networks}. In recent times, social media platforms have emerged as an effective medium for commercial houses, and their objective is to maximize profit. In this paper, we study the Profit Maximization in Closed Social Network (PMCSN) Problem in the context of viral marketing. The input to the problem is a closed social network and two positive integers $\ell$ and $B$. This problem asks to choose nodes as the seed nodes within a given budget $B$, and during the entire diffusion process, each node is restricted to choose a maximum of $\ell$ links for information diffusion, and the objective is to maximize the profit earned by the seed set. PMCSN problem is a generalization of the Influence Maximization problem which is shown to be a NP-hard problem. We propose two solution approaches for PMCSN problem. The first one is a sampling based approximate solution, and the second one is marginal gain-based heuristic solution. Sample complexity, running time, and the space requirement of the proposed approach have been analyzed. A set of experiments have been conducted with real-world, publicly available social network datasets. From the experiments, we observe that the seed set and the diffusion links chosen by the proposed approach led to more profit compared to the baseline methods. The whole implementation and data are available at: https://github.com/PoonamSharma-PY/ClosedNetwork.
\end{abstract}
\section{Introduction}
\emph{Social Networks} have emerged as an effective medium for promoting and branding the products introduced by some commercial houses \cite{chen2025algorithmic}. To do this, a commercial house chooses a limited number of highly influential users and they become the initial adopters of the product. Now, the hope is that some of the initial adopters will like the item and share good words about the product with their neighbors. Some of the neighbors will be influenced and buy the item. This process will go on, and at the end of the diffusion process, a large number of users of the network will be influenced. This method is called \emph{Viral Marketing}, where the goal is to maximize the profit. The key computational problem arises in this context is that given a social network and a positive integer $k$, the goal is to choose $k$ many influential users such that the initial activation of these nodes leads to maximum profit \cite{bhattacharya2019viral,tang2017profit,khan2016revenue,lu2012profit,chalermsook2015social,zhang2016profit,zhu2017maximizing}. There is significant literature on this topic and the root of all these works goes down to the seminal work on Influence Maximization in Social Networks by Kempe et al. \cite{kempe2003maximizing}. Due to the practical applicability of the problem, it has been studied extensively in different variants, and a large number of methodologies are available. For a comprehensive survey, readers are advised to refer to \cite{banerjee2020survey}.
\par One of the important phenomena of social networks relevant to the profit maximization problem is the \emph{diffusion of information} and due to this information, innovation, ideas, concepts, rumors, etc., spread from one node to the next one \cite{bakshy2012role}. To study the diffusion process in a social network, several models have been introduced and studied. One of them is the Independent Cascade Model, which captures the independent behavior of the users \cite{wang2012scalable}. In this model, every user in the network is in the `non-influenced' state, and a small subset of the users are made to be in the `influenced' state externally. Now, the information will be diffused from the influenced nodes in the discrete timesteps with the following rule. An influenced user at timestep $t$ will make a single try to influence each of their non-influenced neighbors, and the success probability will be as mentioned in the corresponding edge probability. The diffusion process ends when no more node activation is possible. Most of the existing studies on information diffusion are through all the links. However, this may not always be useful, as described in the following case study, which brings the motivation of this study.
\par In many practical situations, it is meaningful to limit the diffusion to a small number of outgoing links. Consider the following scenario where an academic institute is trying to promote a conference. Consider that the diffusion starts with the faculty members of the institute. Naturally, there will be many non-academic persons in the contact (or friend, follower, etc.) list. In our context, it may not be useful to diffuse the information to non-academic persons. Social networks where for every user it is possible to restrict the diffusion process within a limited number of neighbors are called \emph{Closed Social Networks} \cite{choi2017trust,huang2022influence}. In closed networks, the diffusion of every user is limited to a subset of its neighbors. As of date, there exist many closed social networks such as WhatsApp, WeChat, etc. However, to the best of our knowledge, profit maximization has not been studied yet on closed social networks. The PMCSN problem extends the classical Influence Maximization problem, which has been proven to be NP-hard \cite{kempe2003maximizing}. In this paper, we bridge this gap by formulating the profit maximization problem in closed social networks, and in particular, we make the following contributions:
\begin{itemize}
    \item We study the profit maximization problem in the closed social network environment, and this is the first study in this direction.
   \item  We propose a sampling-based approximate solution approach and a marginal gain-based heuristic solution with detailed analysis.
    \item A set of experiments has been conducted to highlight the effectiveness and efficiency of the proposed solution approaches. 
\end{itemize}
\par The rest of the paper has been organized as follows. Section \ref{Sec:BPD} describes the background information and defines the problem formally. Section \ref{Sec:Solution} describes the proposed solution approaches. Section \ref{Sec:Experiments} describes the experimental evaluation of the proposed solution approach. Finally, Section \ref{Sec:Conclusion} concludes our study and gives future research directions. 
\section{Background and Problem Definition} \label{Sec:BPD}
In this section, we describe the background of our problem, and we define the problem formally. Initially, we describe Closed Social Networks.
\subsection{ (Closed) Social Networks}
A social network is modeled by a directed weighted graph $G = (V, E, P)$ where $V(G)$ is the set of vertices and $E(G)$ is the set of edges of the graph $G$. The vertex set $V(G) = \{u_1, u_2,..., u_n\}$ represents users of the network. The edge set $E(G)$ of the graph represents the social ties among the users, e.g., $({u_i}{u_j})$ is an edge in $G$ if $u_i$ and $u_j$ have a social relation. $n$ and $m$ denote the number of nodes and edges of the network, respectively. There is an influence probability corresponding to each edge, i.e., $P: E(G) \longrightarrow (0,1]$. The neighbors of a user are connected to it by the edge. The set of outgoing and incoming neighbors of the node $u_i$ is defined as $N_{out}(u)$ and $N_{in}(u)$ and defined as $N_{out}(u_i)=\{u_j: (u_iu_j) \in E(G) \}$ and $N_{in}(u_i)=\{u_j: (u_ju_i) \in E(G) \}$, respectively. For any edge $(u_iu_j)$, the influence probability of the edge is denoted by $P_{u_i \rightarrow u_j}$.  $V_{< \ell}$ and $V_{\geq \ell}$ denote the set of nodes having out-degree strictly less than $\ell$ and greater than or equal to $\ell$, respectively.   
\par In our study, we consider a special type of social network called a \textit{closed social network}. A closed social network is a subset of a larger social network in which connections are limited to a specific group of individuals, based on shared attributes, interests, or associations.  
We select at most $\ell$ outgoing edges for each online user such that the profit earned from the seed set nodes via these selected edges is maximized. Therefore, the network formed is called a diffusion network.
Now, we state the Diffusion Network of a given network in Definition \ref{Def1:lSubnetwork}.
\begin{definition}[Diffusion Network] \label{Def1:lSubnetwork}
    Given a social network $G = (V, E, P)$ and a positive integer $\ell$, a diffusion network of $G$ is denoted by $G_{D}$ and defined as a graph $G_{D}(V, E^{'})$ where the vertex set of $G_{D}$ is the same as the vertex set of $G$ and every vertex of $G_D$ has out-degree at most $\ell$, and hence, $E^{'} \subseteq E$.
\end{definition}
\subsection{Diffusion in Social Networks}
In this study, we assume that the diffusion of information in the underlying network is happening through the Independent Cascade Model, and this has been described in Definition \ref{Def:ICM}.
\begin{definition} [Independent Cascade Model] \label{Def:ICM}
    In this model, diffusion happens in discrete time steps starting from an initially active set. A node can either be in an active (i.e., influenced) or inactive (i.e., non-influenced) state. Every inactive node at timestep $t$, will get a single chance to activate its inactive neighbor and will succeed with the probability mentioned as edge weight. The diffusion process stops when no more node activation is possible.
\end{definition}
At the end of the diffusion process, the set of nodes that are in the active state is called the set of influenced nodes. Given a seed set $S$, $I(S)$ denotes the set of nodes that are influenced. Next, we define the influence of a given seed set in Definition \ref{Def:Influence}.
\begin{definition} [Influence of a Seed Set] \label{Def:Influence}
    Given a social network $G(V,E,P)$, and a seed set $S$, the influence of $S$ is quantified using the influence function $\sigma$, which maps each possible subset of users to its expected influence, i.e., $\sigma:2^{V} \longrightarrow \mathbb{R}_{0}^{+} $ where $\sigma(\emptyset)=0$.
\end{definition}
\subsection{Profit Maximization in Social Networks}
We consider that every user of the network is associated with a cost and a benefit value and they are formalized as the cost function $C$ and benefit function $b$, respectively, i.e., $C: V(G) \longrightarrow \mathbb{R}^{+}$ and $b: V(G) \longrightarrow \mathbb{R}^{+}$. For any user $u \in V$, its associated cost and benefit values are denoted by $C(u)$ and $b(u)$, respectively. The selection cost signifies the amount of incentive that needs to be paid to him if he is selected as a seed user. The associated benefit value with a user signifies the amount of incentive that can be earned from the user if he is influenced. For a subset of seed nodes $S \subseteq V$, its selection cost is denoted by $C(S)$ and defined as $C(S)=\underset{u \in S}{\sum} \ C(u)$. Now, we define the notion of benefit earned by a seed set in a closed social network in Definition \ref{Def3:SeedSetBenefit}.

\begin{definition}[Benefit earned by the Seed Set] \label{Def3:SeedSetBenefit}
   Given a closed social network $G(V,E,P)$ along with the cost and benefit function $C$ and $b$, and a seed set $S$, and a positive integer $\ell$, the earned benefit by $S$ in one of the diffusion networks $G_{D} \in \alpha(G)$, is denoted by $\beta_{G_{D}} (S)$ and defined as the sum of the benefit values of the influenced nodes in $G_{D}$. Let $I_{G_{D}}(S,u)$ be a boolean variable whose value is $1$, if $u$ is influenced by the seed set $S$ and $0$ otherwise. This can be represented by the following conditional equation:
\[
    I_{G_{D}}(S,u)= 
\begin{cases}
    1,& \text{if } u \text{ is influenced by } S \\
    0,              & \text{otherwise}
\end{cases}
\]
   
Hence, 
\begin{equation}
    \beta_{G_{D}} (S)= \underset{u \in V(G_{D})}{\sum} \ I_{G_{D}}(S,u) \cdot b(u)
\end{equation}
The benefit earned by the seed set $S$ in the given closed social network $G(V,E,P)$ is denoted by $\beta(S)$ and defined as the expected value of the benefit earned across all possible diffusion networks. Mathematically, this has been represented in Equation \ref{Eq:Benefit}.

\begin{equation} \label{Eq:Benefit}
    \beta(S)=  \underset{G_{D} \in \alpha(G)}{\sum} Pr(G_{D}) \cdot  \beta_{G_{D}} (S)
\end{equation}
Here, $Pr(G_{D})$ is the probability that the diffusion network $G_{D}$ is generated, which is equal to $\frac{1}{|\alpha(G)|}$.
\end{definition}

\begin{definition}[Profit earned by the Seed Set] \label{Def4:SeedSetProfit}
    Given a closed social network $G = (V, E, P)$, a positive integer $\ell$, and a seed set $S \subseteq V$, then the profit earned by the seed set $S$ is denoted by $\phi(S)$ and defined as the difference between the benefit earned by the seed set and the cost of seed set.
    \begin{equation}\label{Eq3:Profit}
        \phi(S) = \beta(S) - C(S)
    \end{equation}
\end{definition}
Now, we formally state the Profit Maximization in Closed Social Networks (PMCSN) Problem, which is defined in Definition \ref{Def5:PMCSN}.
\begin{definition}[Profit Maximization in Closed Social Network (PMCSN Problem)] \label{Def5:PMCSN}
    Given a close social network $G = (V, E, P)$ along with the cost and benefit functions $C$ and $b$, and two positive integers $\ell$ and $B$, this problem aims at finding the following:
    \begin{itemize}
    \item for every user $u \in V_{\geq \ell}$, choose $\ell$ outgoing edges to construct the diffusion network.
    \item a subset of users $S \subseteq V(G)$ within the Budget $B$, i.e., $\underset{u \in S}{\sum} C(u) \leq B$,
        
    \end{itemize}
Hence, the PMSCN Problem can be modeled as a discrete optimization problem stated in Equation \ref{Eq:PMCSN}. 
\begin{equation}\label{Eq:PMCSN}
          (S^{*}, E^{*} )\longleftarrow \underset{\substack{S \subseteq V(G) \text{ and } C(S) \leq B \\ |E_{u}| =\ell }}{argmax} \ \phi(S)
     \end{equation}
\end{definition}

\section{Proposed Solution Approach} \label{Sec:Solution}
In this section, we describe two proposed solution approaches. The first subsection contains a sampling-based approach and the next subsection contains 
\subsection{Sampling-Based Approach}

\begin{algorithm}[h]
\caption{Sampling Based Approximate Approach for the PMCSN Problem}
\label{Algorithm1:Sampling}
\KwData{Closed Social Network $G(V,E,P)$, two positive integers $B$, and $\ell$}
\KwResult{A subset $S \subseteq V$ such that $C(S) <= B$ and at most $\ell$ many outgoing links for each $u \in V$}
$\alpha(G) \gets \text{Generate all possible diffusion networks}$\;
$\mathcal{A} \longleftarrow \text{Consider a subset of diffusion networks}$\;
$\phi_{max} \longleftarrow -1$, $G_{max} \longleftarrow \emptyset$, $S \gets \emptyset$\;
\ForEach{$\mathcal{G} \in \mathcal{A}$}{
    $S_{\mathcal{G}} \gets \text{Seed Set of diffusion network } \mathcal{G} $\;
   $\phi_{\mathcal{G}} \longleftarrow \text{Calculate the profit from seed nodes }S_{\mathcal{G}}$\;
    \If{$\phi(S_{\mathcal{G}}) > \phi_{max}$}{
        $\phi_{max} \gets \phi(S_{\mathcal{G}})$, $G_{max} \gets \mathcal{G}$,  $S \gets S_{\mathcal{G}}$\;
        
    }
}
\Return $S, \ \phi_{max}, \ G_{max} $
\end{algorithm}

\paragraph{\textbf{Complexity Analysis}}
Let, $x$ be the sample size. As mentioned previously, in each diffusion network, there can be at most $ \frac{B}{C_{min}}$ many seed nodes. In each diffusion network, computing profit using the IC Model will take $\mathcal{O}(\frac{B}{C_{min}} \cdot (m+n))$ time. As the sample size is $x$, the time requirement for the sampling-based solution approach will be of $\mathcal{O}(\frac{B}{C_{min}} \cdot (m+n) \cdot x)$. The space requirement depends upon the sampling strategy. If we do the online sampling, then in each iteration of the \texttt{for} loop, we will create a new diffusion network and compute the profit over there. In this case, the space requirement will be to store one diffusion network, which is of $\mathcal{O}(m+n)$. In case of offline sampling, all the required samples need to be generated before profit computation. Hence, in this case the space requirement will be $\mathcal{O}(x \cdot (m+n))$. 
\begin{theorem}
    In the worst case, the time and space requirements of the sampling-based approach will be of $\mathcal{O}(\frac{B}{C_{min}} \cdot (m+n) \cdot x)$ and $\mathcal{O}(x \cdot (m+n))$, respectively, where $x$ is the sample size.
\end{theorem}
\paragraph{\textbf{Sample Bound Estimation}} Here, we give a calculation on the number of samples required such that the difference between the maximum profit using optimal seed set and diffusion links and the best one in the sample set will be small with high probability. However, before that, we explain some simple facts. The maximum profit that can be earned by a seed set with any set of diffusion links is the sum of the benefit values of all the nodes minus the minimum selection cost among all the nodes. We denote this by $\mathcal{B}$, i.e., $\mathcal{B}= \underset{u \in V}{\sum} b(u) - C_{min}$. We consider that the minimum profit earned by any seed set is $0$. Hence, for any randomly sampled diffusion network, the computed profit will be within the range $0$ and $\mathcal{B}$. Hence, for any randomly sampled diffusion network, the earned profit will be in between $0$ and $\mathcal{B}$. We denote the ratio between the estimated profit and the maximum profit earned by the symbol $\rho$. 

\begin{theorem}
For any given $\epsilon, \delta \in (0,1)$ if the sample size is more than $\frac{ ln(\frac{2}{\delta})}{2 \cdot \epsilon^{2} \cdot \rho^{2}}$ then the probability that the difference between the actual and estimated profit is less than $\epsilon$ will be at least $1-\delta$.
\end{theorem}

\subsection{Marginal Gain Based Heuristic (HEU) Solution}
In this section, we describe an efficient heuristic solution for our problem. In this proposed approach, for every node $v \in V_{\geq \ell}$, we choose $\ell$ many high-degree nodes, and this constitutes the diffusion network. We denote this diffusion network by $G_{D}(V,E^{'})$. Now, from this diffusion network, we select the seed nodes in the following way. First, we sample out $\frac{n}{k} \log \frac{1}{\epsilon}$ many non-seed nodes, and from these sampled nodes, we select nodes based on maximum marginal influence gain. Observe that sampling is done in every iteration. Algorithm \ref{Algorithm2:Heuristic} describes this procedure in the form of pseudocode.
\begin{algorithm}[t]
\caption{Marginal Gain Based Heuristic Solution Approach for the PMCSN Problem}
\label{Algorithm2:Heuristic}
\KwData{Social Network $G(V,E,P)$, two positive integers $B$, and $\ell$}
\KwResult{A subset $S \subseteq V$ such that $C(S) <= B$ and at most $\ell$ many outgoing links for each $u \in V$}
$\text{Initialize } \mathcal{D}[1 \ldots n] \longleftarrow 0$\;
$\text{Initialize } V(G_{D}) \longleftarrow V(G) \text{ and } E(G_{D}) \longleftarrow \phi$\;
\For{$i=1 \text{ to }n$}{
$\mathcal{D}[i] \longleftarrow \text{Compute the degree of } v_i$\;
}
\For{$\text{Each } v \in V(G)$}{
$N_{\geq \ell}(v) \longleftarrow \text{ Select } \ell \text{ many high degree neighboring nodes}$\;
\For{$\text{Each } u \in N_{\geq \ell}(v)$}{
$E(G_{D}) \longleftarrow E(G_{D}) \cup \{(vu)\}$\;
}
}
$\text{Initialize } S \longleftarrow \phi$ and $\epsilon \longleftarrow 0.1$ \;
\While{$B > 0$}{
$\text{Set } k \longleftarrow \frac{B}{C_{min}}$\;
$S^{'} \longleftarrow \text{Sample } \frac{n}{k} \log \frac{1}{\epsilon} \text{ many nodes from } V(G_{D}) \setminus S$\;
$v^{*} \longleftarrow \underset{v \in S^{'} \text{ and }C(v) \leq B}{argmax} \  \frac{\sigma(S \cup \{v\}) \ - \ \sigma(S)}{C(v)}$\;
$S \longleftarrow S \cup \{v^{*}\}$\;   
}
$\phi_{max} \longleftarrow \text{Compute Profit in } G_{D}$\;
\Return $S, \ \phi_{max}, \ G_{max} $
\end{algorithm}
\par Now, we analyze Algorithm \ref{Algorithm2:Heuristic} to understand its time and space requirements. Initialization steps will take $\mathcal{O}(1)$ time. Computing the degrees of the nodes will take $\mathcal{O}(n^{2})$ time. We sort the vertices based on their degree, which takes $\mathcal{O}(n\log n)$ time. Now, for every node to mark $\ell$ many highest neighbouring nodes from the sorted list will take $\mathcal{O}(n)$ time. Hence, the time requirement to execute from Line No. $5$ to $8$ will take $\mathcal{O}(n^{2})$ time. It is important to estimate how many times the \texttt{while} loop from line $10$ to $13$ will be executed. Assume that $C_{min}$ denotes the minimum selection cost of the nodes, i.e., $C_{min}= \underset{u \in V(G_{D})}{min} C(u)$. In the worst case $\mathcal{O}(\frac{B}{C_{min}})$ many nodes can be selected. Now, within the \texttt{while} loop, first we set $k$ to $\frac{B}{C_{min}}$. Next, to sample $\frac{n}{k} \log \frac{1}{\epsilon}$ many elements from the set $V(G_{D}) \setminus S$ will take $\mathcal{O}(\frac{n}{k} \log \frac{1}{\epsilon})$ time. Computing marginal gain in one iteration will take $\mathcal{O}(n \cdot \frac{n}{k} \log \frac{1}{\epsilon} \cdot (m+n))$. Finding the node with the largest maximum profit gain will take $\mathcal{O}(1)$ time. Hence, the time requirement to execute from Line $10$ to $14$ will be of $\mathcal{O}(n^{2} \cdot (m+n) \log \frac{1}{\epsilon})$. Hence, the total running time of Algorithm \ref{Algorithm2:Heuristic} will be of $\mathcal{O}(n^{2} \cdot (m+n)  \log \frac{1}{\epsilon})$. Extra space consumed by Algorithm \ref{Algorithm2:Heuristic} to store the following data structures:
\begin{itemize}
    \item To store the degrees of the nodes which will take $\mathcal{O}(n)$ space (Line $4$).
    \item To store the $\ell$ many random nodes from the neighborhood of a node which will take $\mathcal{O}(\ell)$ space. (Line No. $6$)  
    \item  To store the edges of the diffusion network, which will take $\mathcal{O}(n^{2})$ space in the worst case. (Line No. $8$)  
    \item To store the set of nodes present in $S^{'}$ which will take $\mathcal{O}(\frac{n}{k} \cdot \log \frac{1}{\epsilon})$ and this will reduce to $\mathcal{O}(n)$. (Line No. $12$).
    \item To store the marginal profit gain to cost ratio of every node which will take $\mathcal{O}(\frac{n}{k} \cdot \log \frac{1}{\epsilon}))$ space and this will reduce to $\mathcal{O}(n)$. (Line No. $13$).
\end{itemize}
As in the worst case $\ell$ can be of $\mathcal{O}(n)$, it can be observed that the space complexity of Algorithm \ref{Algorithm2:Heuristic} will be of $\mathcal{O}(n^{2})$. Hence, Theorem \ref{Th:3} holds.
\begin{theorem} \label{Th:3}
    The time and space complexity of Algorithm \ref{Algorithm2:Heuristic} will be of $\mathcal{O}(n^{2} \cdot (m+n)  \log \frac{1}{\epsilon})$ and $\mathcal{O}(n^{2})$, respectively.
\end{theorem}

\section{Experimental Evaluation} \label{Sec:Experiments}

\paragraph{\textbf{Dataset Description}}
In our experiments, we have used two real-world datasets,
listed in Table \ref{Table: Table1Dataset} showing their basic statistics. Here, $d_{max}$ and $d_{avg}$ denote the maximum and average degree of a network. 

\begin{table}[h]
\caption{Basic Statistics of the Datasets}
\label{Table: Table1Dataset}
\centering
\begin{tabular}{|l|l|c|c|c|c|}
\hline
\textbf{Dataset}       & \textbf{Type} & \textbf{Nodes} & \textbf{Edges} & \textbf{$d_{max}$} & \textbf{$d_{avg}$} \\ \hline
\textbf{Email-Eu-core (Euemail)} \cite{10.1145/3097983.3098069} & Directed      & 1005           & 25571          & 347                & 33.25              \\ \hline
\textbf{Facebook} \cite{NIPS2012_7a614fd0}      & Undirected    & 4039           & 88234          & 1045               & 43.7               \\ \hline
\end{tabular}
\end{table}
\vspace{-0.3in}

\paragraph{\textbf{Experimental Setup and and Goals}}
In this study, there are mainly three parameters whose values need to be fixed for the experimentation. These parameters are: (i) Diffusion Probability, (ii) Budget, and (iii) Value of $\ell$. We consider two probability settings, namely \emph{trivalency} and \emph{weighted cascade}. In trivalency setting, the influence probability of an edge is chosen uniformly at random from the set $\{0.1, 0.01, 0.001\}$. In weighted cascade setup, for every edge $(u_iu_j)$, its influence probability will be $\frac{1}{deg(u_i)}$. We have experimented with the $5$ different budget values: $500$, $1000$, $1500$, $2000$, and $2500$. We have experimented with $4$ different $\ell$ values: $4$, $12$, $20$, and $28$. We have compared the performance of our proposed approach with two methods, namely Random and High Degree. In the `Random' algorithm, we randomly select seed nodes within budget, and the $\ell$ outgoing edges per node are also selected randomly to generate the diffusion network onto which the diffusion is performed. In the `High Degree' algorithm, we randomly select top high out-degree nodes as seed nodes within budget, and the $\ell$ outgoing edges per node are also selected randomly to generate the diffusion network onto which the diffusion is performed.

In this study, we have addressed the following research questions: 
\begin{itemize}
    \item \textbf{Impact of Budget}: If the budget is increased, then how does the earned profit change?
    \item \textbf{Impact of the $\ell$}: If the value of $\ell$ is increased, then how does the earned profit change?
    \item \textbf{Impact on Computational Time}: How does the computational time requirement change if the budget and value of $\ell$ change?
\end{itemize}
The experiments were run on a Linux system with an Intel i9, 32 cores at 3.2 GHz, and 64 GiB RAM. All algorithms were implemented in Python 3.13.5 using NetworkX 3.5.

\begin{figure}[h]
\centering
\begin{tabular}{cccc}
\begin{tabular}{c}
\includegraphics[width=0.2199\textwidth]{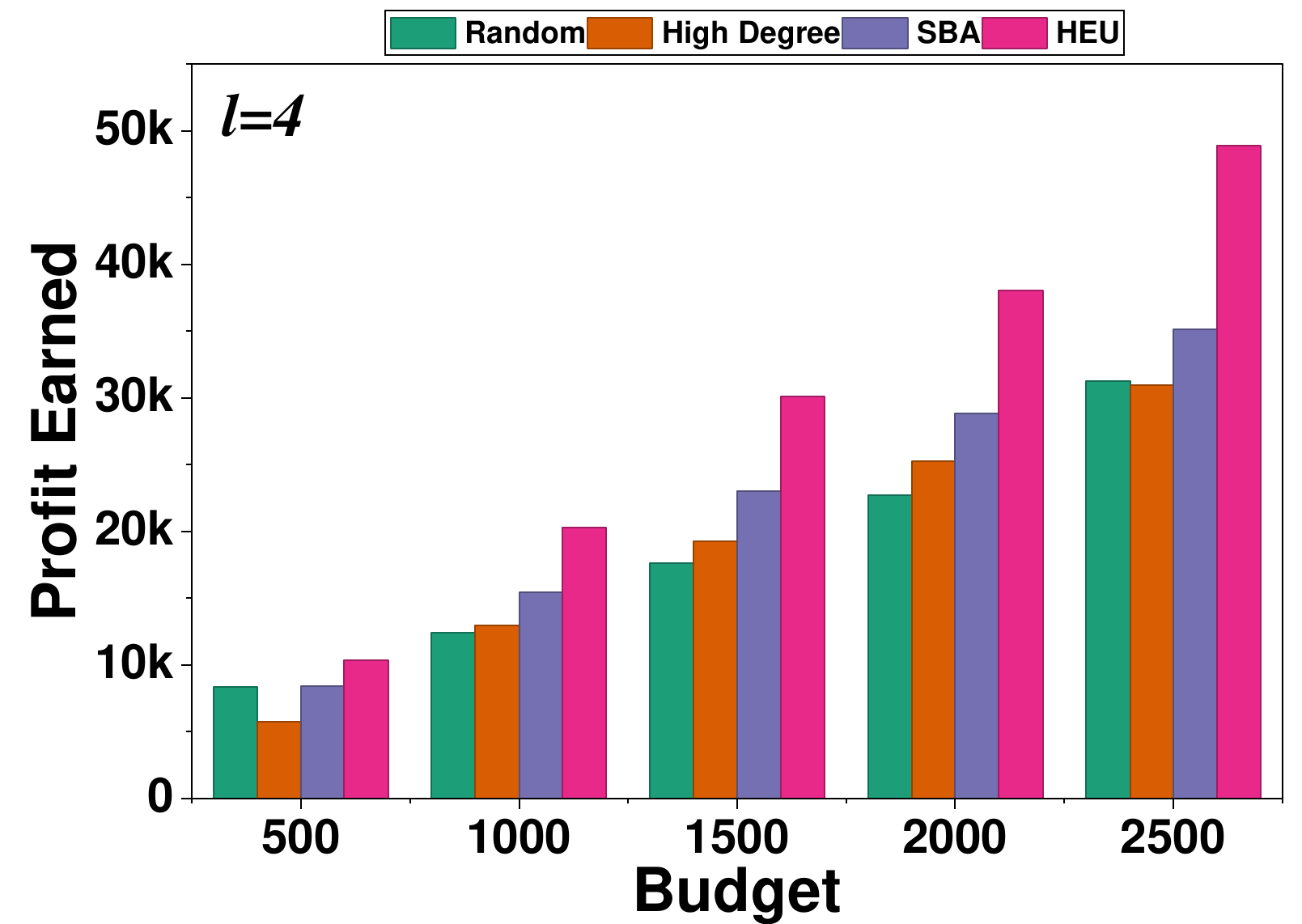} \\
(a) Trivalency
\end{tabular} &
\begin{tabular}{c}
\includegraphics[width=0.2199\textwidth]{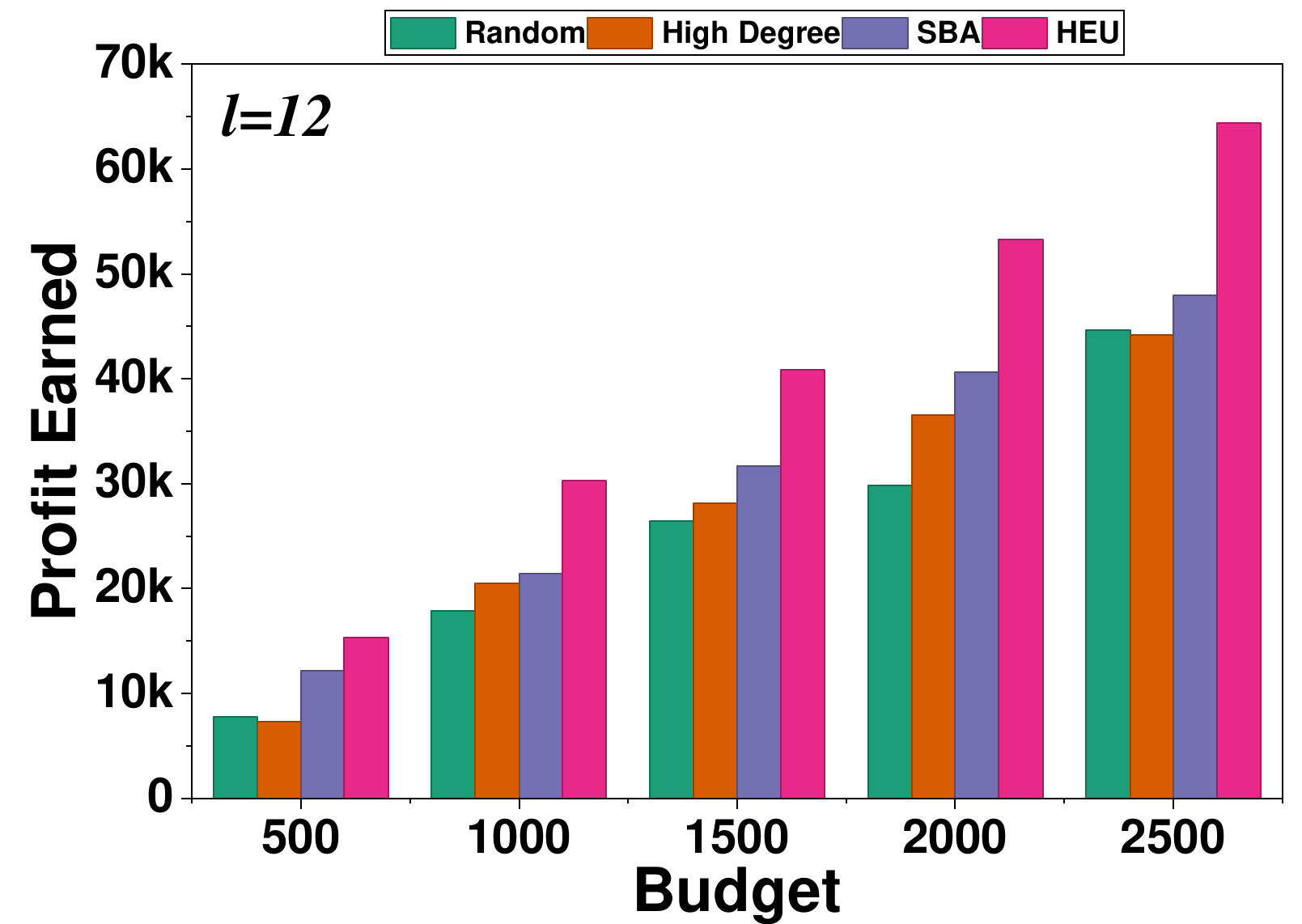} \\
(b) Trivalency
\end{tabular} &
\begin{tabular}{c}
\includegraphics[width=0.2199\textwidth]{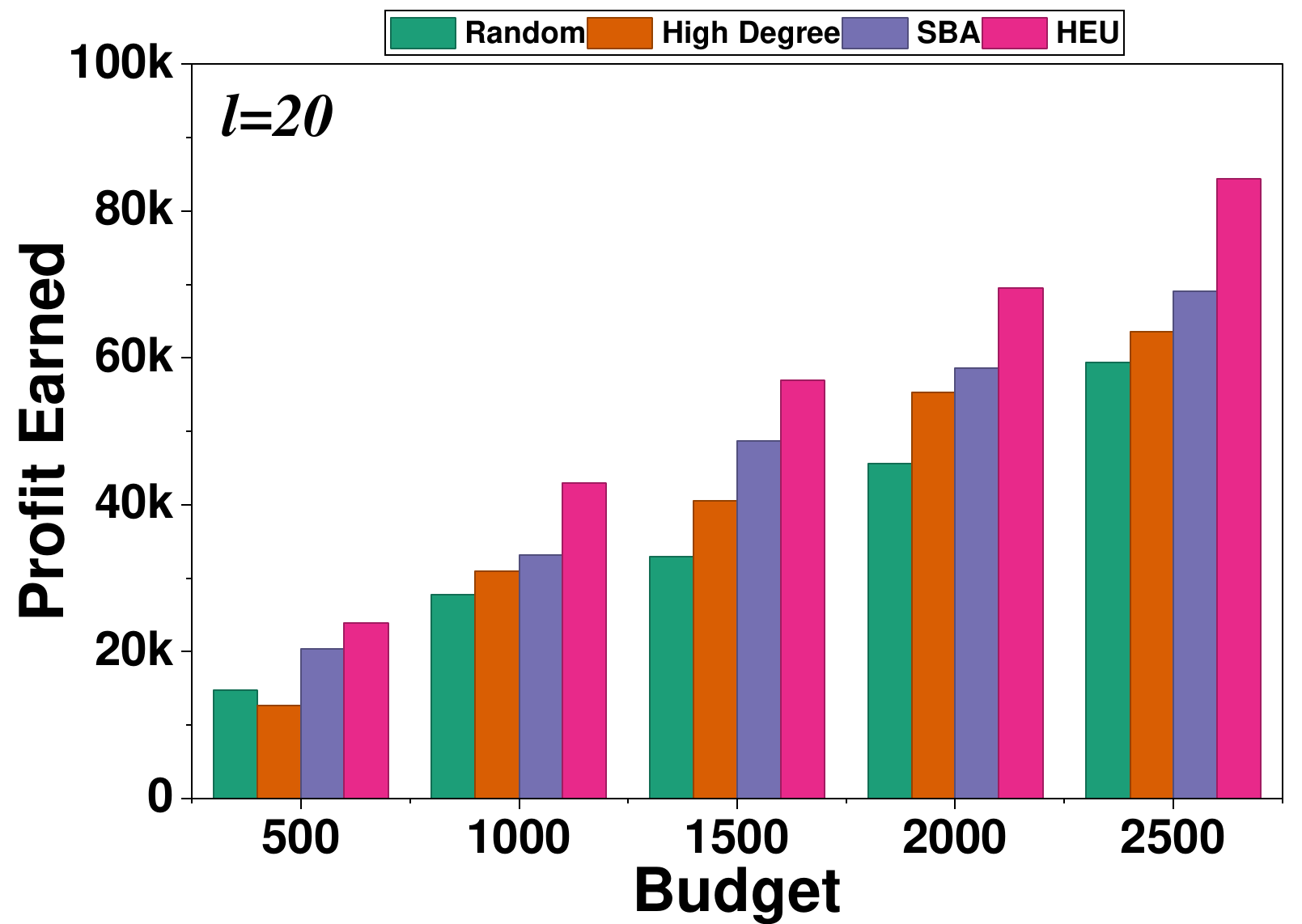} \\
(c) Trivalency
\end{tabular} &
\begin{tabular}{c}
\includegraphics[width=0.2199\textwidth]{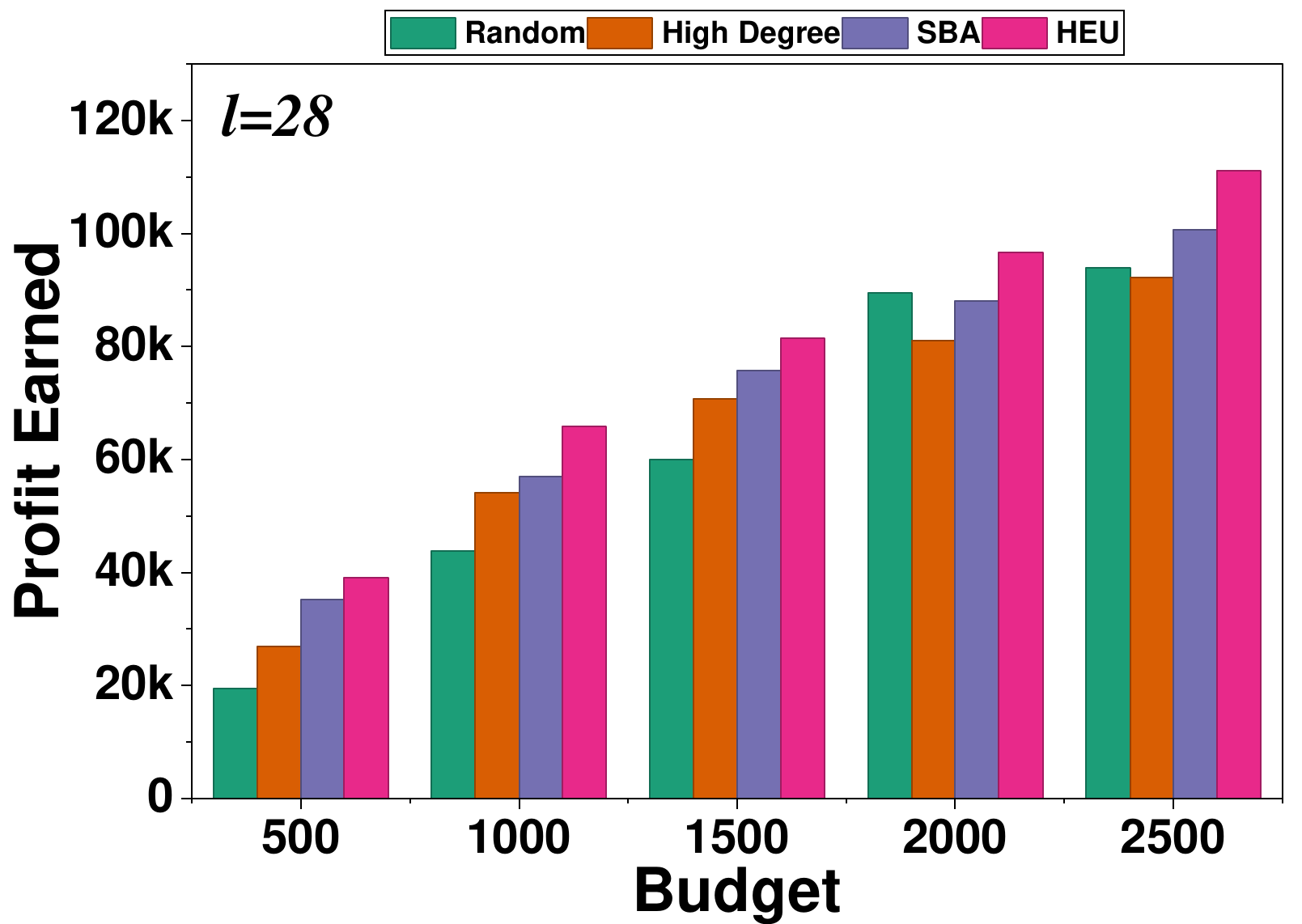} \\
(d) Trivalency
\end{tabular}
\\[6pt]

\begin{tabular}{c}
\includegraphics[width=0.2199\textwidth]{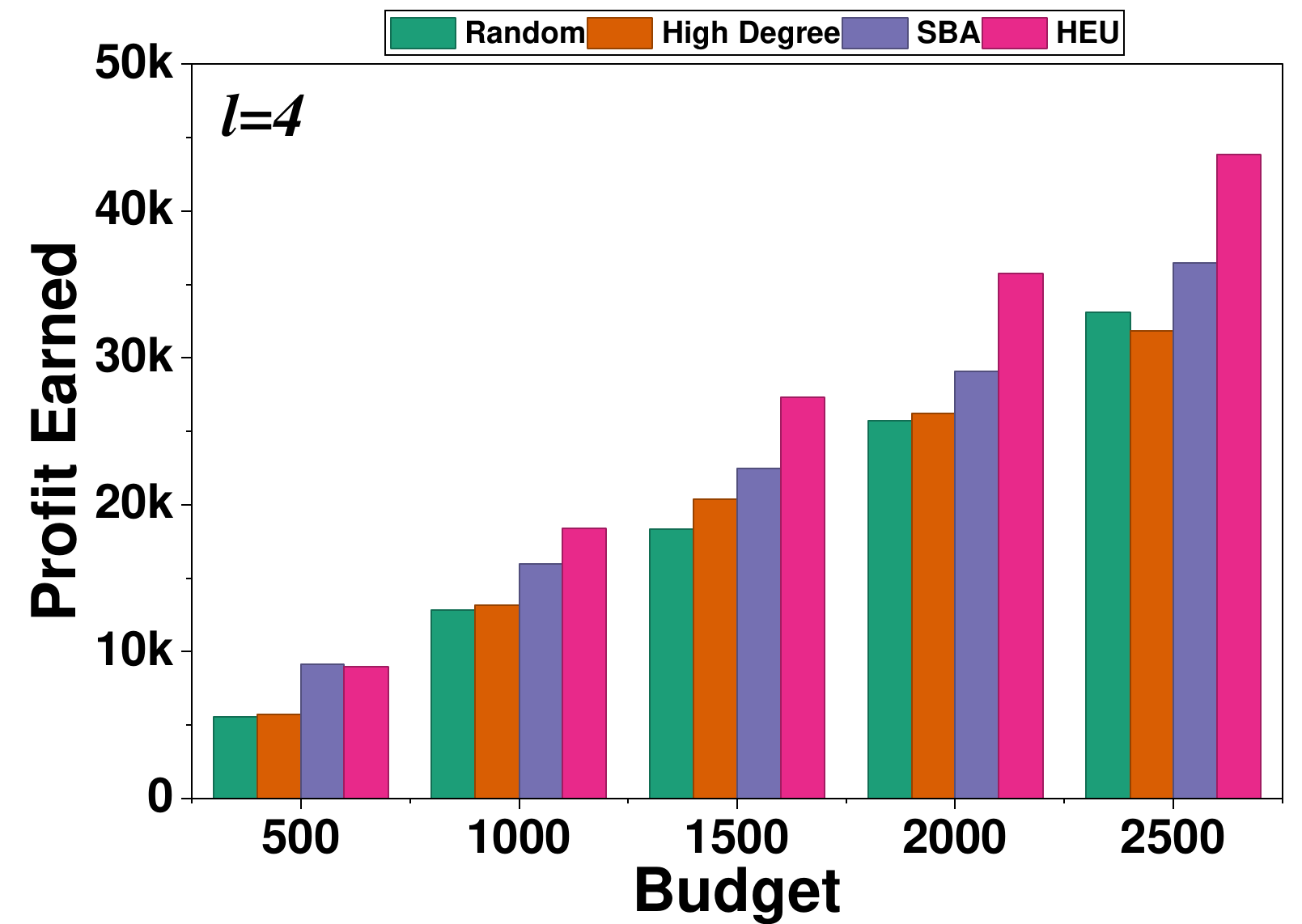} \\
(e) WC
\end{tabular} &
\begin{tabular}{c}
\includegraphics[width=0.2199\textwidth]{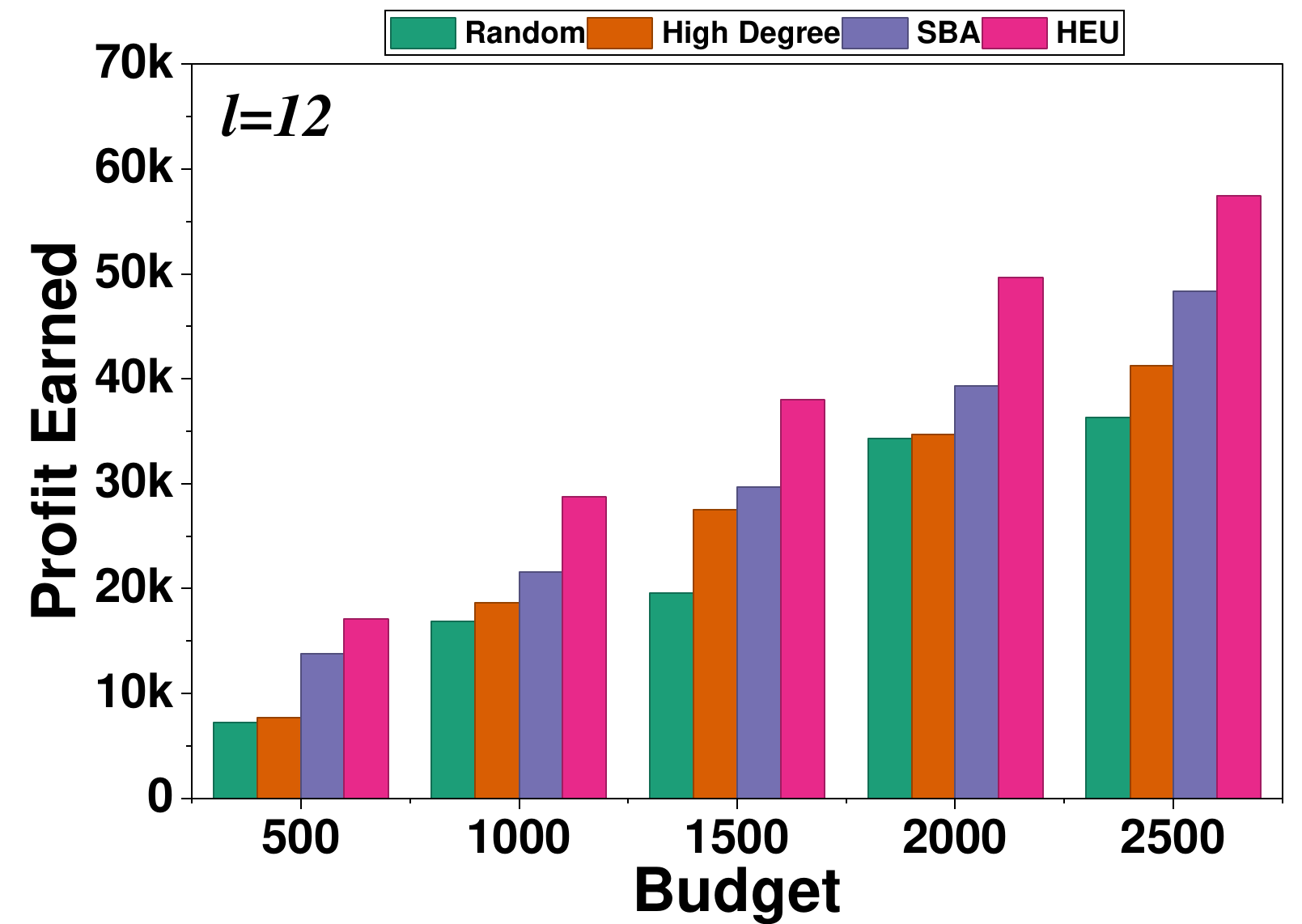} \\
(f) WC
\end{tabular} &
\begin{tabular}{c}
\includegraphics[width=0.2199\textwidth]{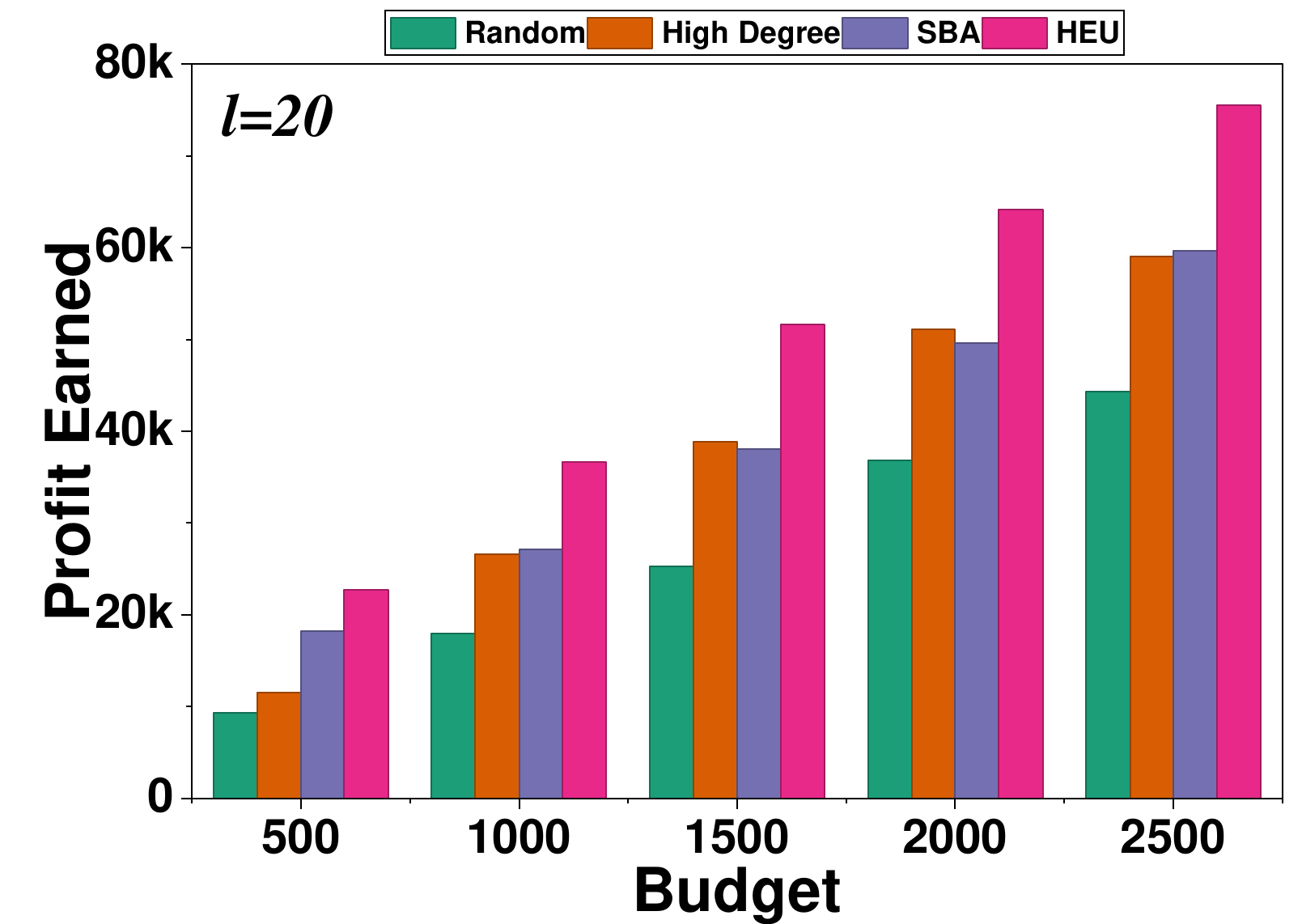} \\
(g) WC
\end{tabular} &
\begin{tabular}{c}
\includegraphics[width=0.2199\textwidth]{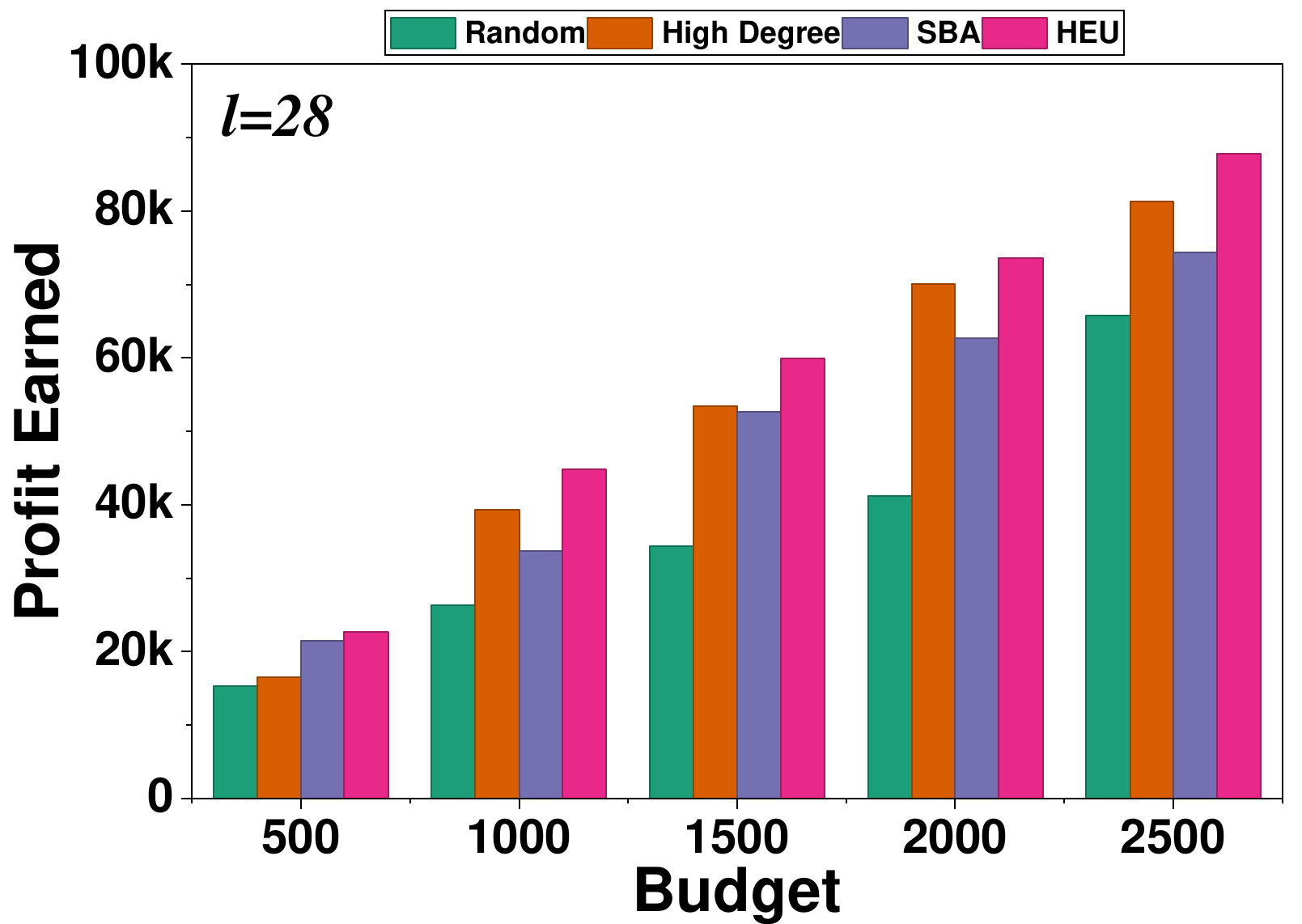} \\
(h) WC
\end{tabular}
\\[6pt]
\begin{tabular}{c}
\includegraphics[width=0.2199\textwidth]{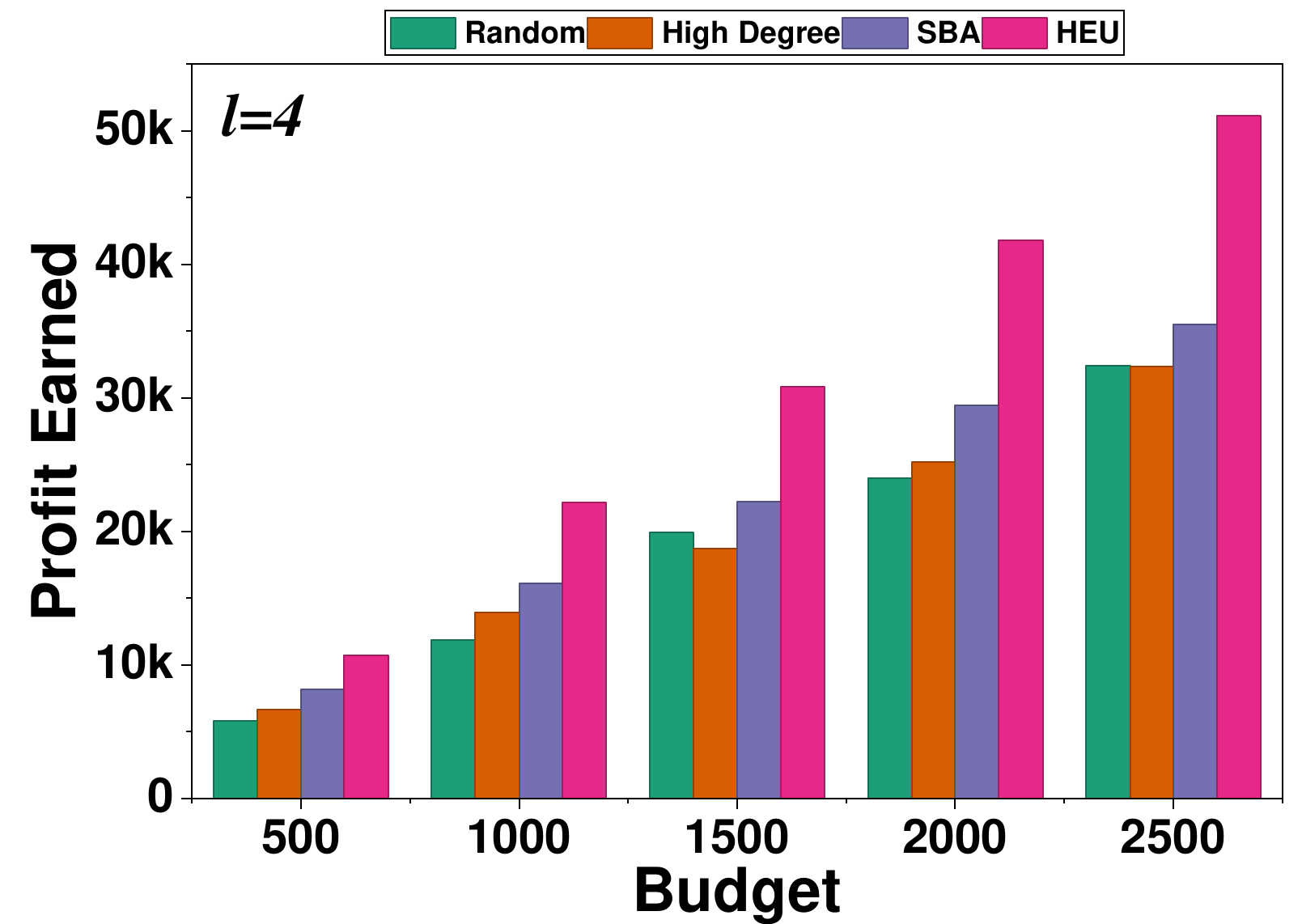} \\
(i) Trivalency
\end{tabular} &
\begin{tabular}{c}
\includegraphics[width=0.2199\textwidth]{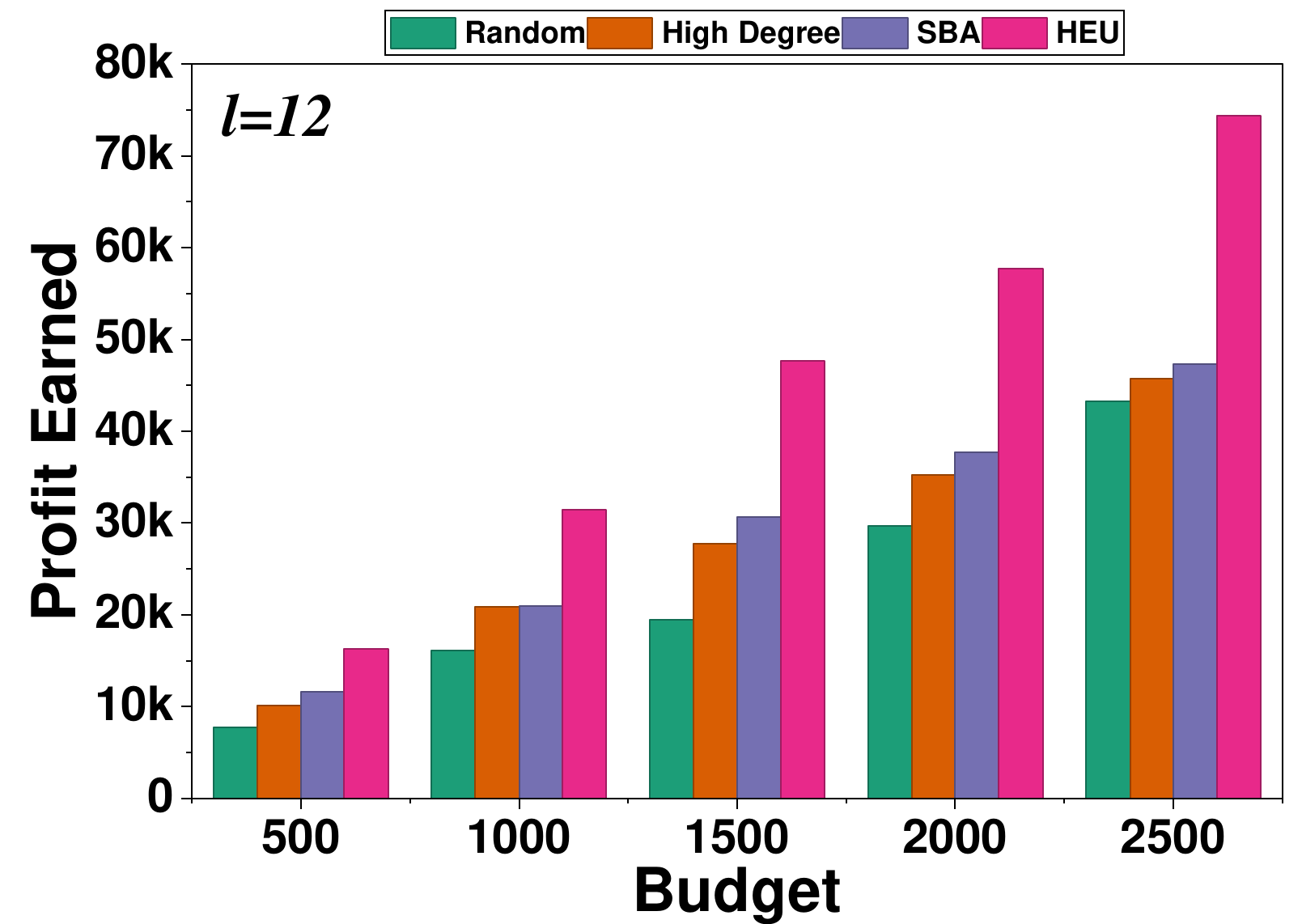} \\
(j) Trivalency
\end{tabular} &
\begin{tabular}{c}
\includegraphics[width=0.219\textwidth]{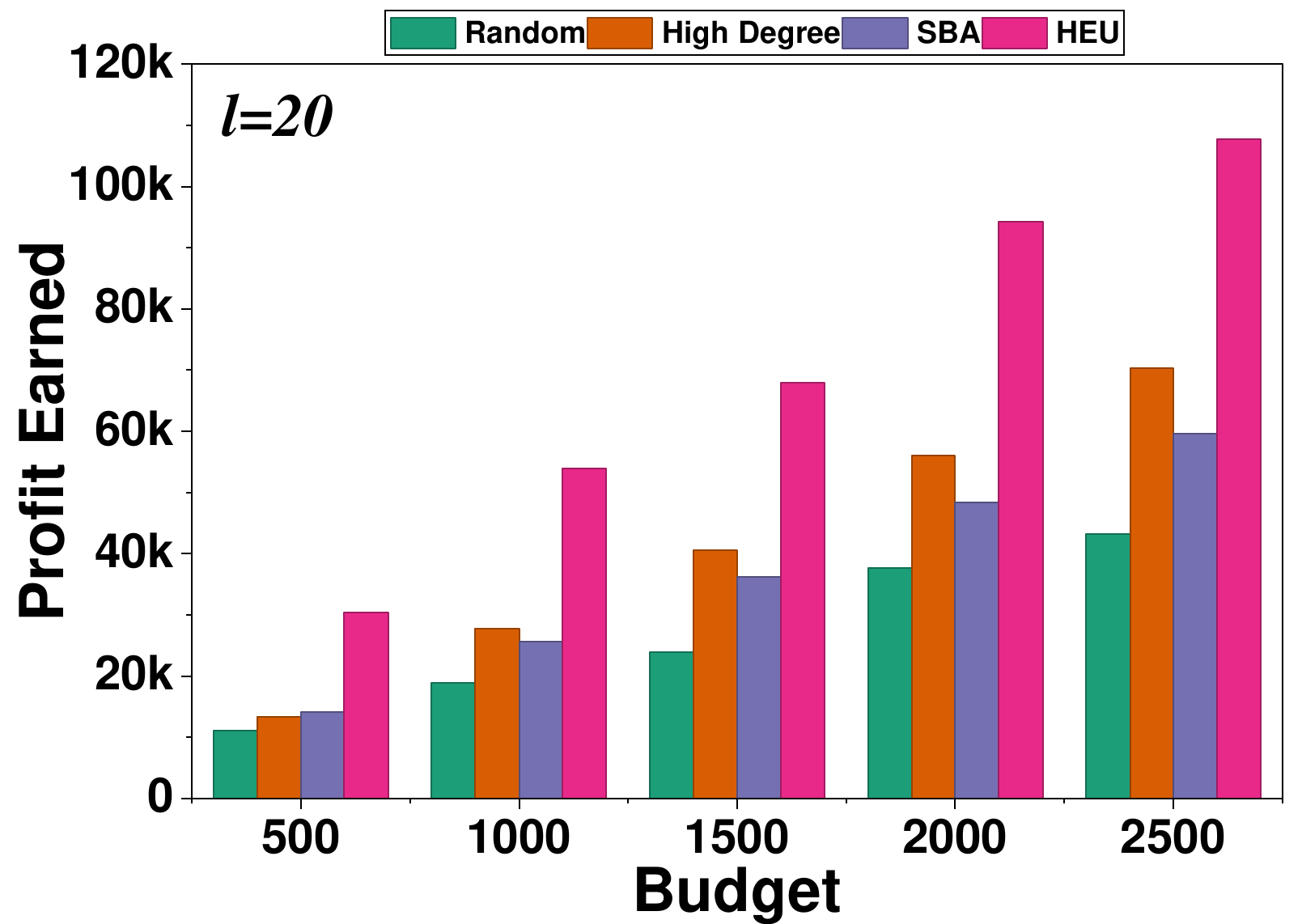} \\
(k) Trivalency
\end{tabular} &
\begin{tabular}{c}
\includegraphics[width=0.2199\textwidth]{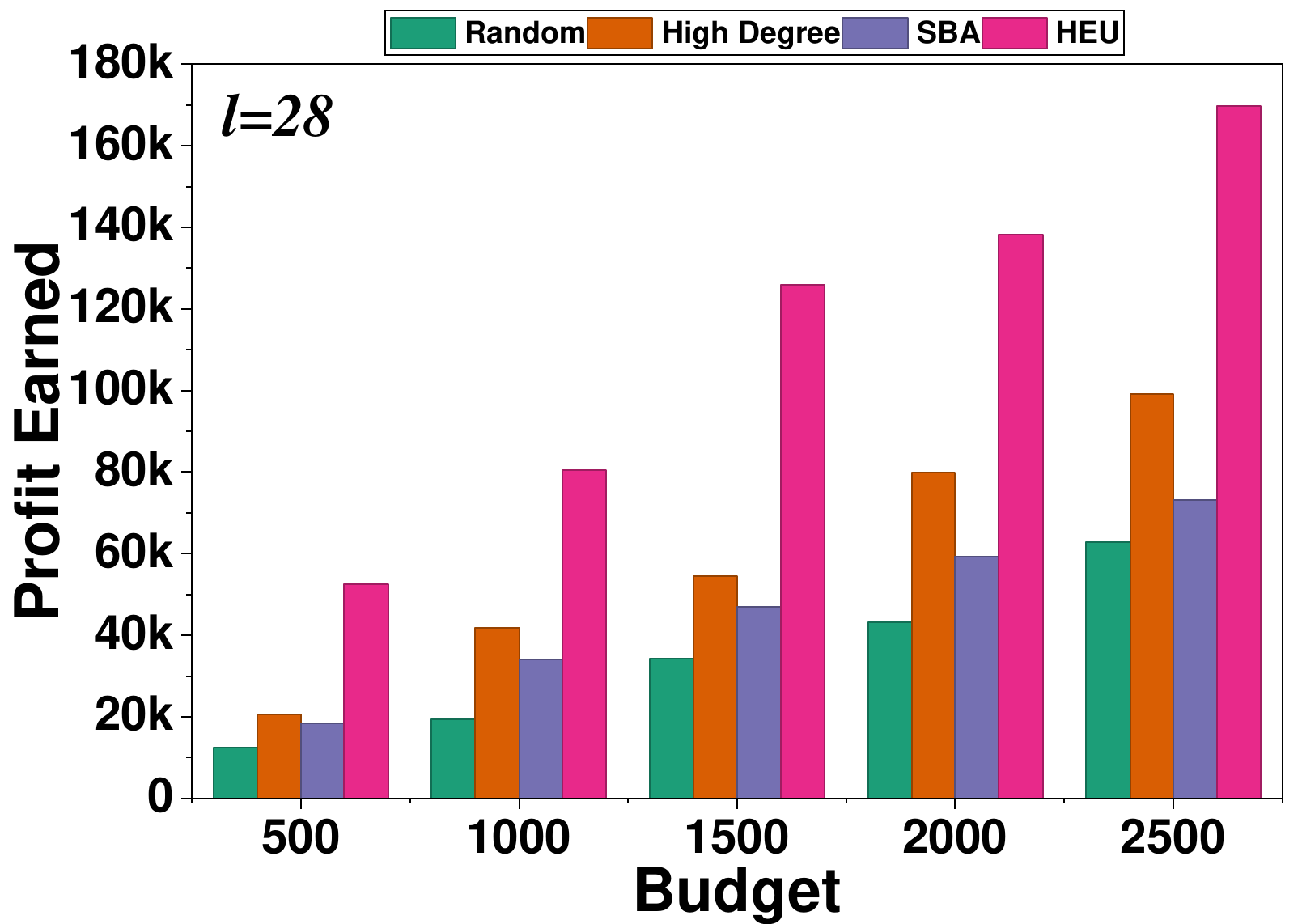} \\
(l) Trivalency
\end{tabular}
\\[6pt]

\begin{tabular}{c}
\includegraphics[width=0.2199\textwidth]{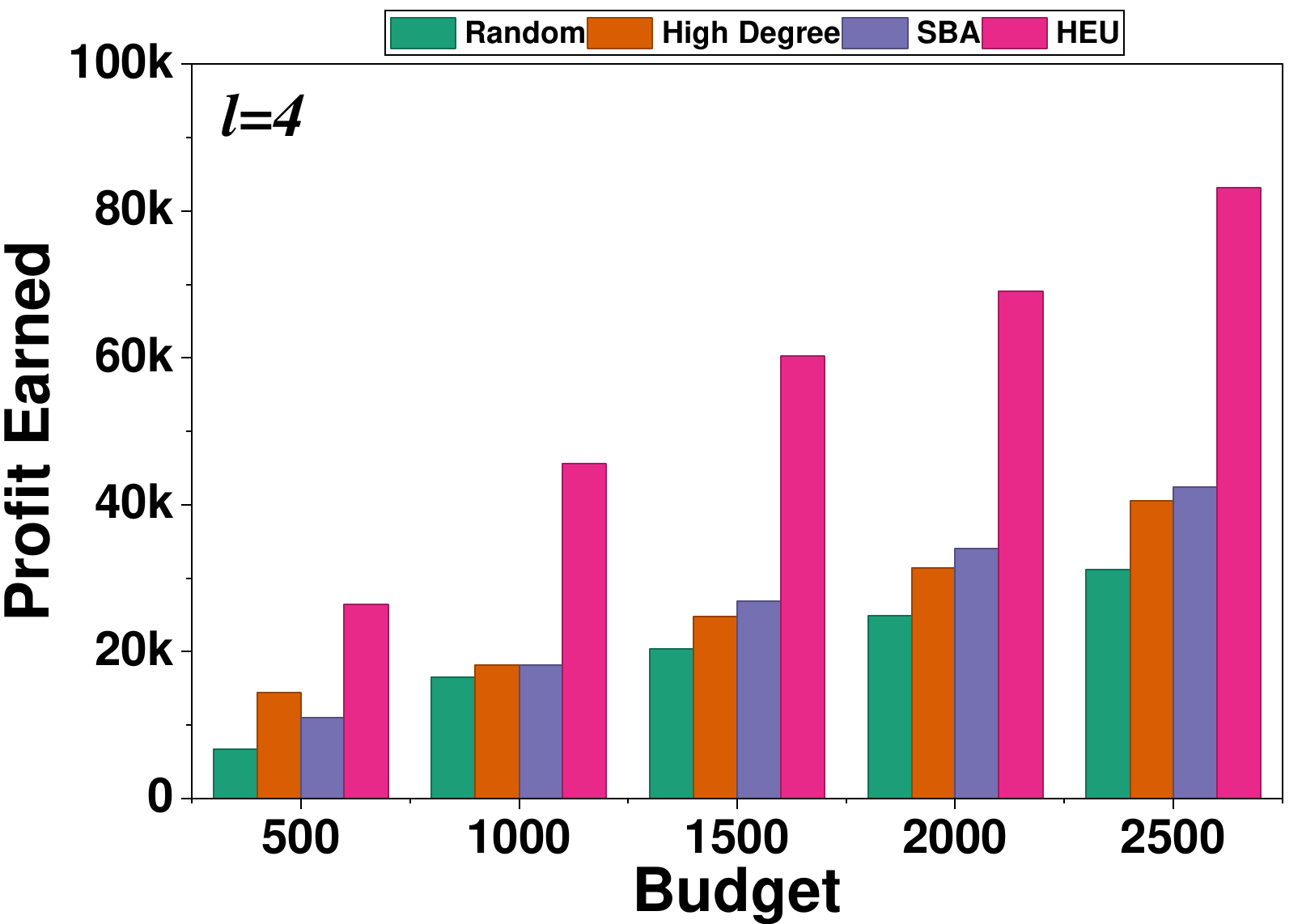} \\
(m) WC
\end{tabular} &
\begin{tabular}{c}
\includegraphics[width=0.2199\textwidth]{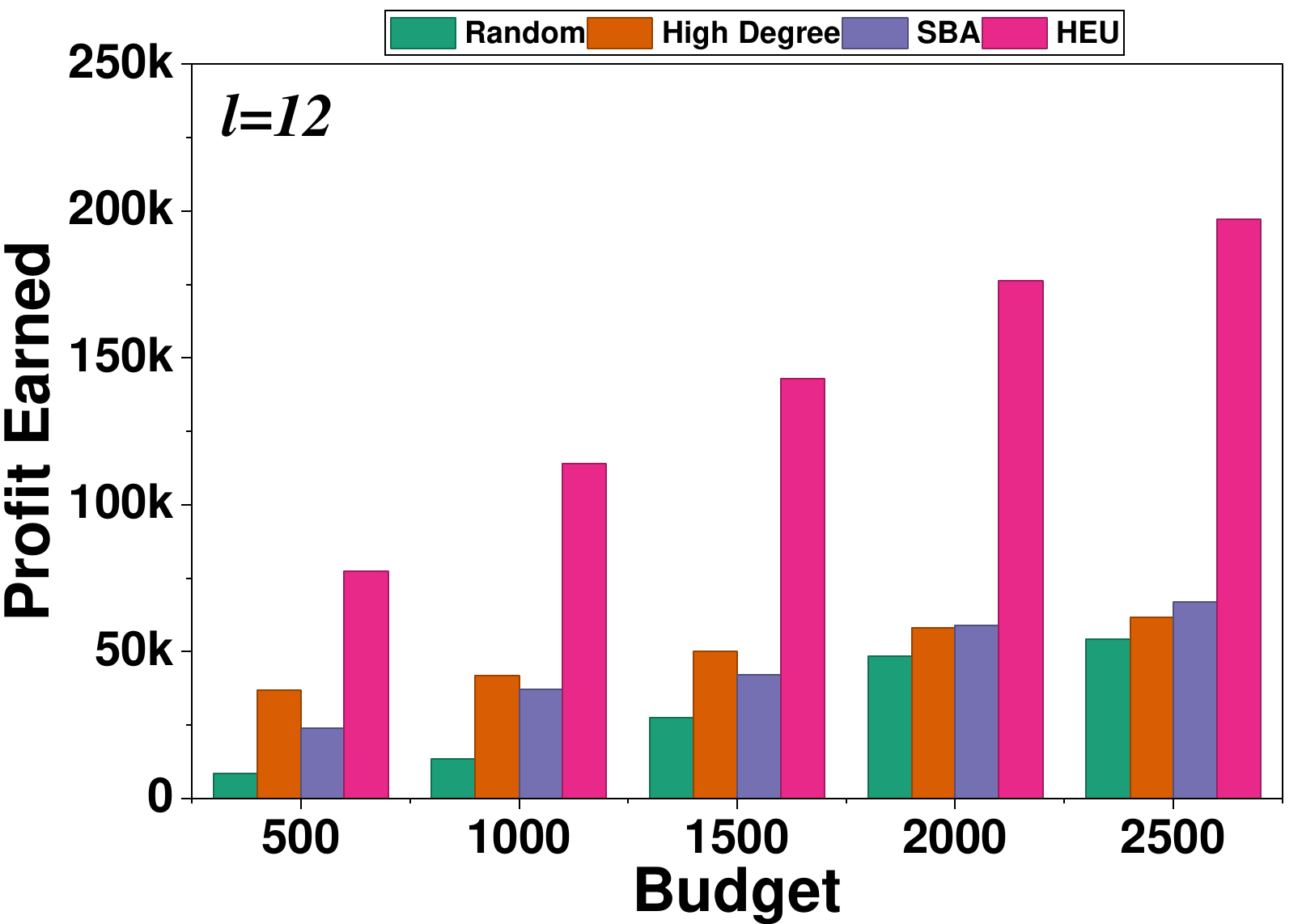} \\
(n) WC
\end{tabular} &
\begin{tabular}{c}
\includegraphics[width=0.2199\textwidth]{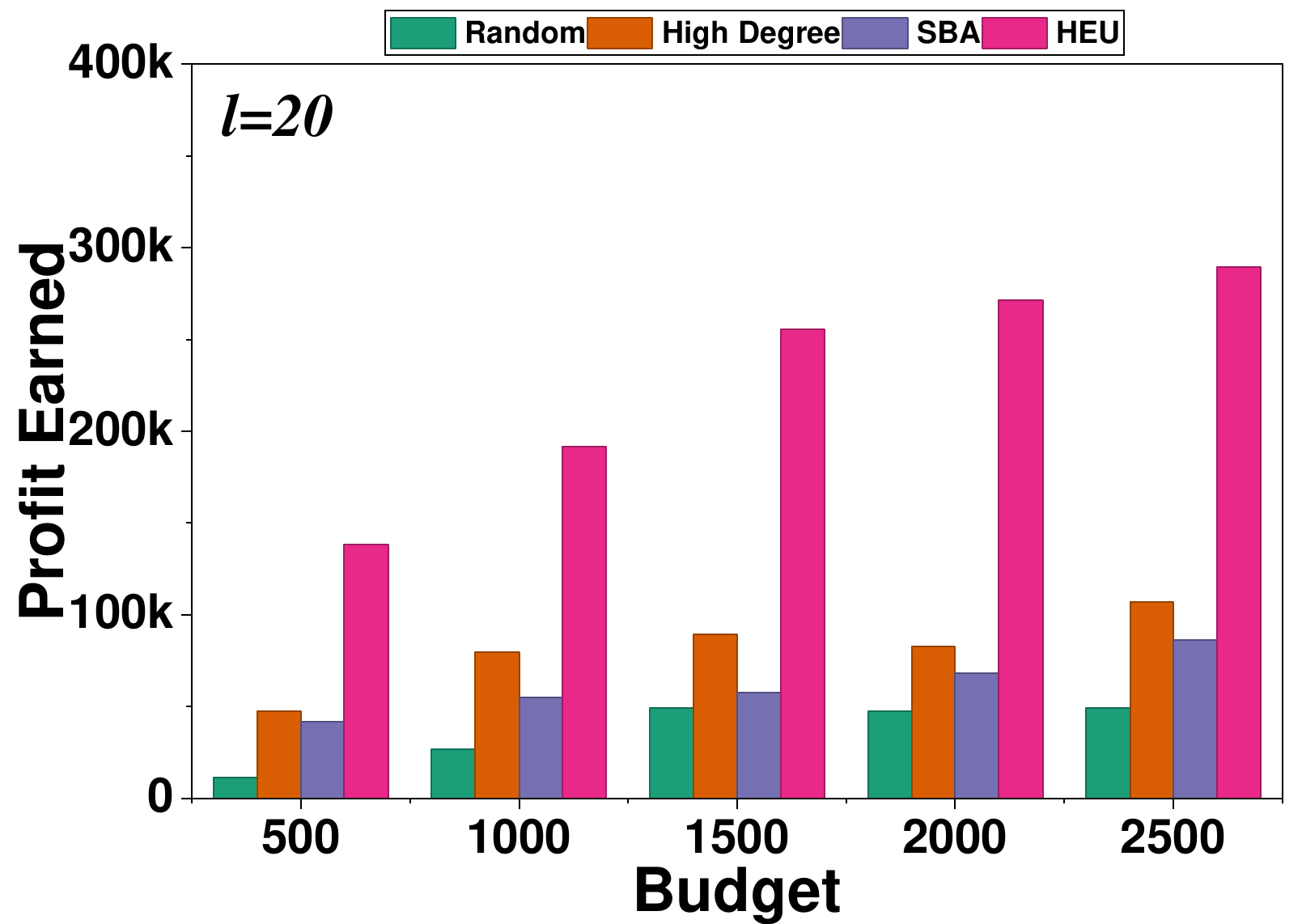} \\
(o) WC
\end{tabular} &
\begin{tabular}{c}
\includegraphics[width=0.2199\textwidth]{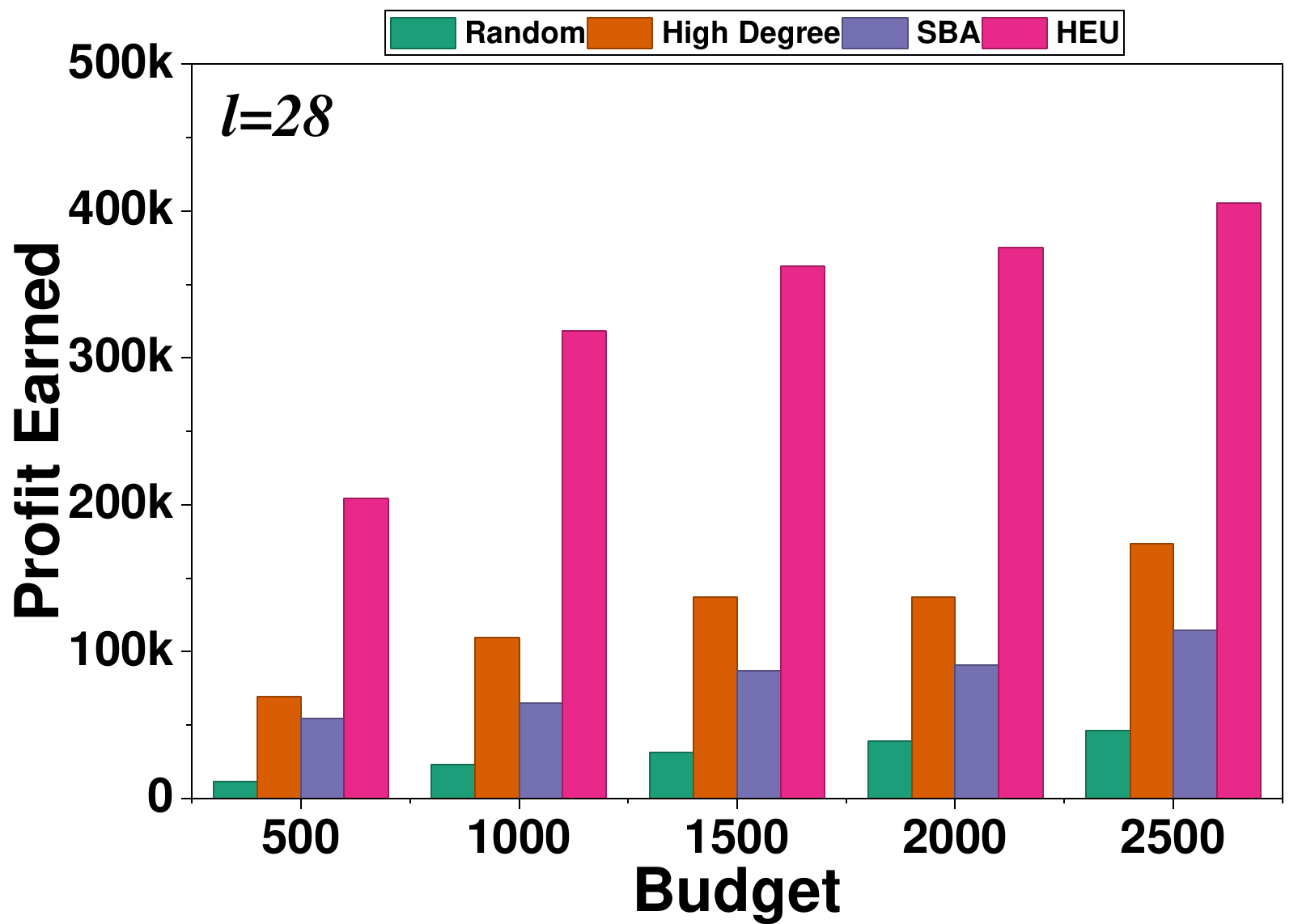} \\
(p) WC
\end{tabular}
\end{tabular}

\caption{\textit{Budget Vs. Profit Earned} Plots for \textit{Euemail} ((a)-(h)) and \textit{Facebook} ((i)-(p)) Dataset}
\label{Plot1:Profit}
\end{figure}

\paragraph{\textbf{Experimental Results and Discussions}}
We now present the experimental results obtained and describe them in detail to address the three research questions posed above.
\paragraph{\textbf{Impact of Budget}}
Initially, we will analyze the outcomes related to the trivalency probability configuration. Fig. $\ref{Plot1:Profit}$(a)-(d) presents graphs of the \textit{Euemail} dataset, illustrating how different algorithms' profits are impacted by increasing budget values. As the budget grows, so does the profit for each algorithm. For example, in Fig. $\ref{Plot1:Profit}$(a), with $\ell = 4$, the baseline algorithm Random produces a profit of $8356$ at a budget of $500$, $12430.62$ at a budget of $1000$, $17633.9$ at a budget of $1500$, $22719.37$ at a budget of $2000$, and $31271.22$ at a budget of $2500$. Similar results can be observed for our methods, SBA and HEU. The \textit{Facebook} dataset also demonstrates a similar trend across all algorithms. The plots are in Fig. $\ref{Plot1:Profit}$(i)-(l). The results for the weighted cascade probability setting are shown in Figures $\ref{Plot1:Profit}$(e)-(h) and $\ref{Plot1:Profit}$(m)-(p) for the \textit{Euemail} and \textit{Facebook} datasets, respectively. The trends of the results of the weighted cascade probability setting are consistent with those of trivalency. For any individual algorithm, as the budget increases, the profit earned also increases.
\paragraph{\textbf{Impact of the $\ell$}}
Increasing $\ell$ (max outgoing edges per node) consistently raises profit. For \textit{Euemail} (Fig. \ref{Plot1:Profit}(a)–(d)), at budget $2000$ and $\ell=4$, High Degree gives $11.11\%$ more profit than Random; at $\ell=12$ it reaches $36542.96$ (a $44\%$ increase over $\ell=4$), and at $\ell=28$ profit rises another $46\%$ over $\ell=20$. With budget $1000$, HEU exceeds SBA by $41\%$ at $\ell=12$, increases $42\%$ at $\ell=20$ over $\ell=12$, and gains another $53\%$ at $\ell=28$ over $\ell=20$. HEU is consistently the highest. For \textit{Facebook} (Fig. \ref{Plot1:Profit}(i)–(l)), at budget $1500$, Random grows from $19902.2$ ($\ell=4$) to $34249.78$ ($\ell=28$). At $\ell=28$, High Degree earns $54406$, SBA $46940.2$, and HEU $125825.5$—a $308\%$ gain for HEU over its $\ell=4$ value. Under Weighted Cascade, larger $\ell$ again boosts profit. For \textit{Euemail} (Fig. \ref{Plot1:Profit}(e)–(h)), at budget $1500$, profits rise from $18336.5$–$27339.5$ ($\ell=4$), to $25306.43$–$51672.31$ ($\ell=20$), and to $34303.3$–$59940.73$ ($\ell=28$). At budget $500$, profits increase from $5547.6$–$9127.81$ ($\ell=4$) to $15249$–$22630.45$ ($\ell=28$); at budget $2500$, from $33074$–$43847$ ($\ell=4$) to $65790.2$–$87798.81$ ($\ell=28$). We observe that Random is lowest, SBA strong, and HEU highest. For \textit{Facebook} Weighted Cascade (Fig. \ref{Plot1:Profit}(m)–(p)), $\ell=4$ yields $6724.5$–$83142.6$, while $\ell=12$ and $\ell=20$ substantially increases all profits. With budget $1500$, profits grow from $20319.45$–$60203.62$ ($\ell=4$) to $49336.53$–$255524.51$ ($\ell=20$). At $\ell=28$, HEU reaches $405468.8$ at budget $2500$, nearly ten times Random. Higher $\ell$ enhances connectivity, spread, and profits, with SBA and especially HEU gaining the greatest advantage.
\begin{figure}[h]
\centering
\begin{tabular}{cccc}
\begin{tabular}{c}
\includegraphics[width=0.2199\textwidth]{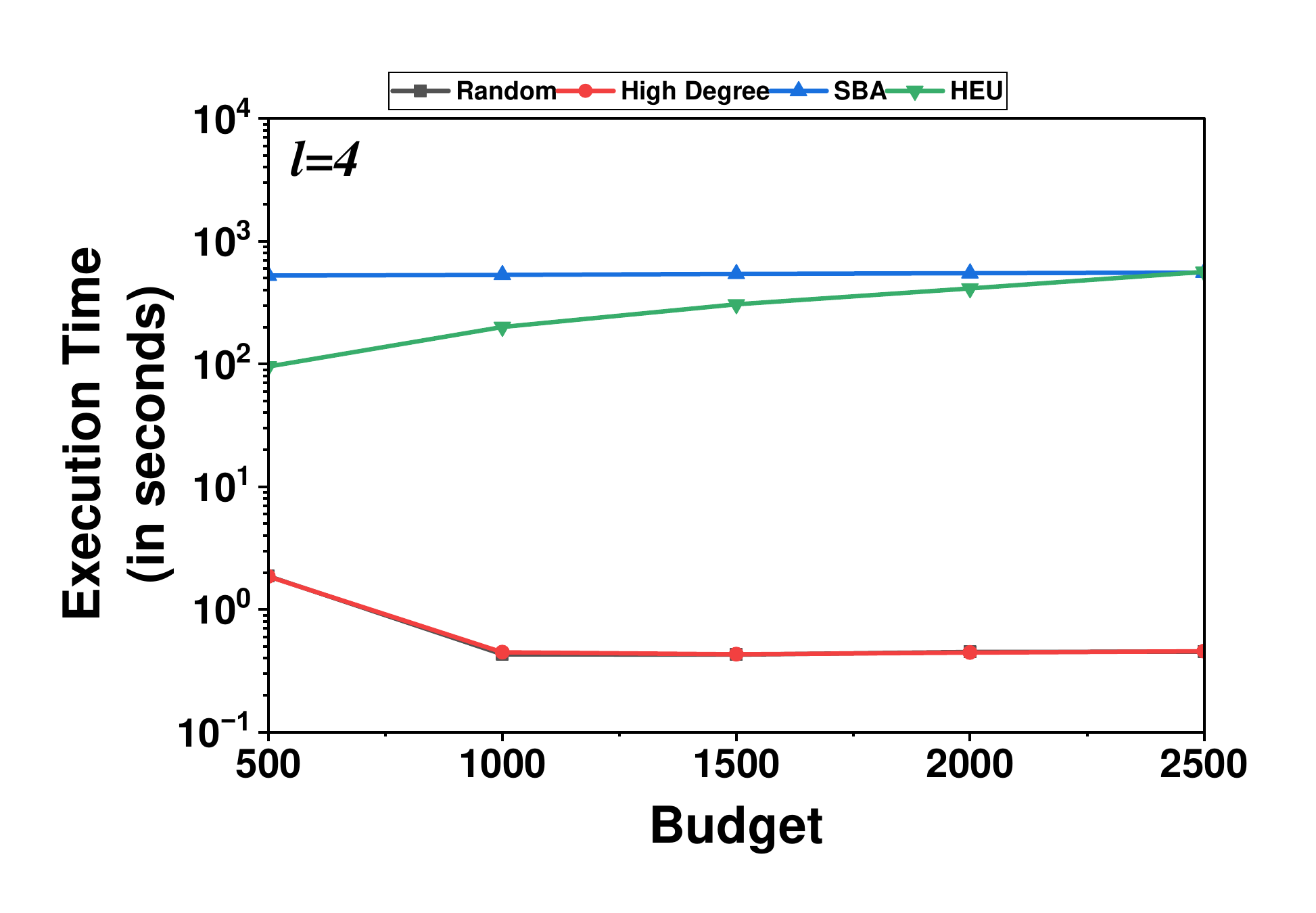} \\
(a) Trivalency
\end{tabular} &
\begin{tabular}{c}
\includegraphics[width=0.2199\textwidth]{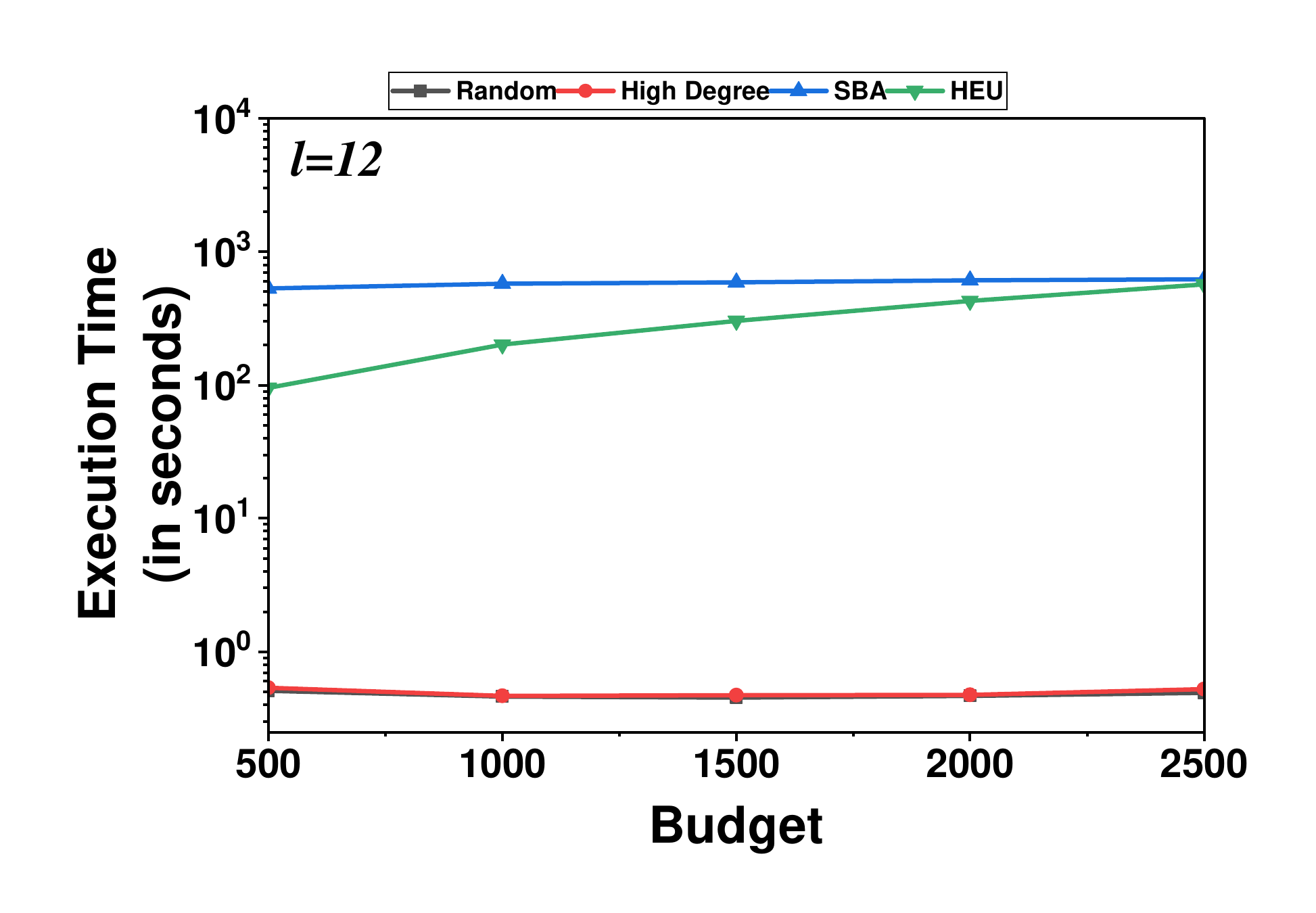} \\
(b) Trivalency
\end{tabular} &
\begin{tabular}{c}
\includegraphics[width=0.2199\textwidth]{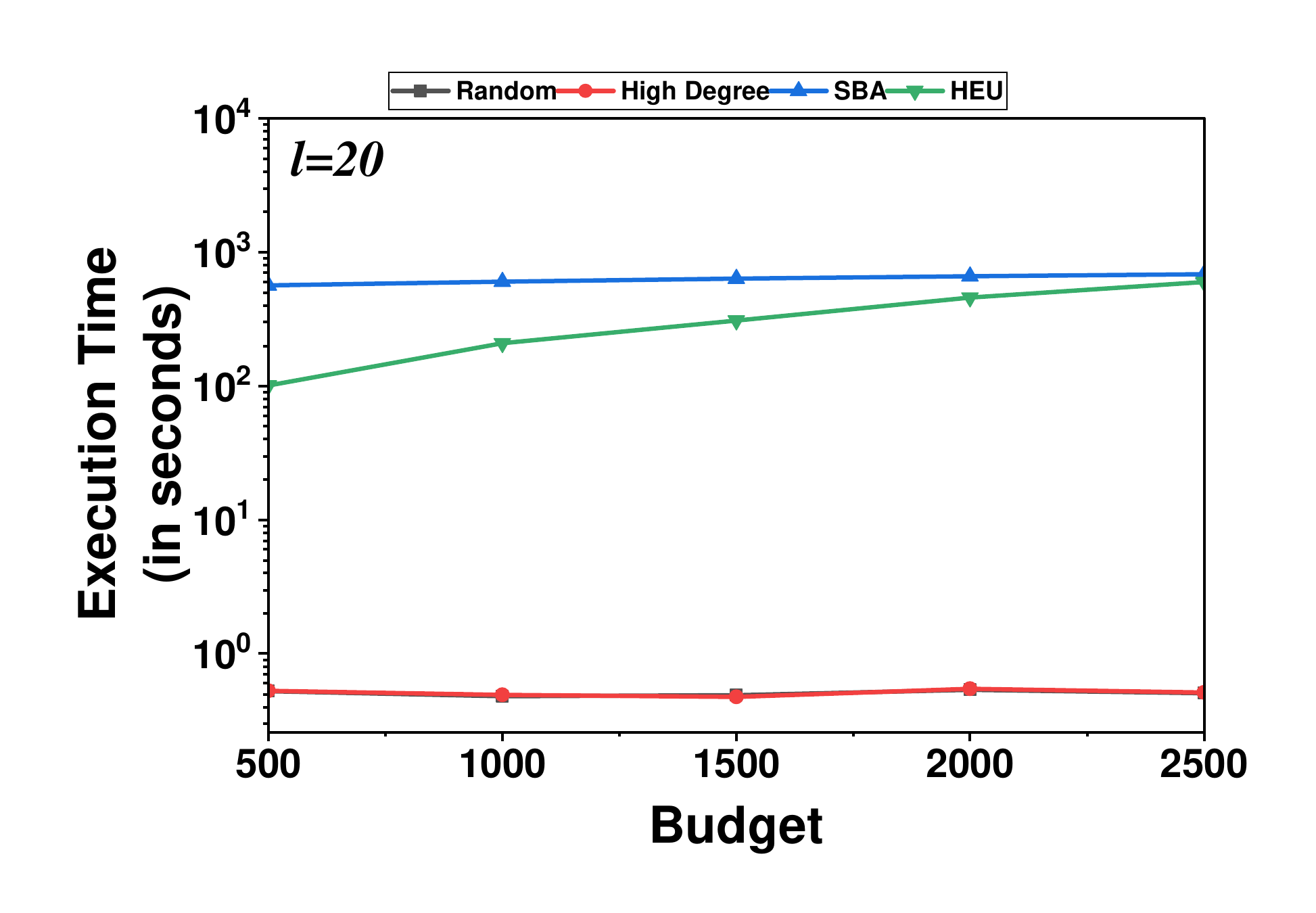} \\
(c) Trivalency
\end{tabular} &
\begin{tabular}{c}
\includegraphics[width=0.2199\textwidth]{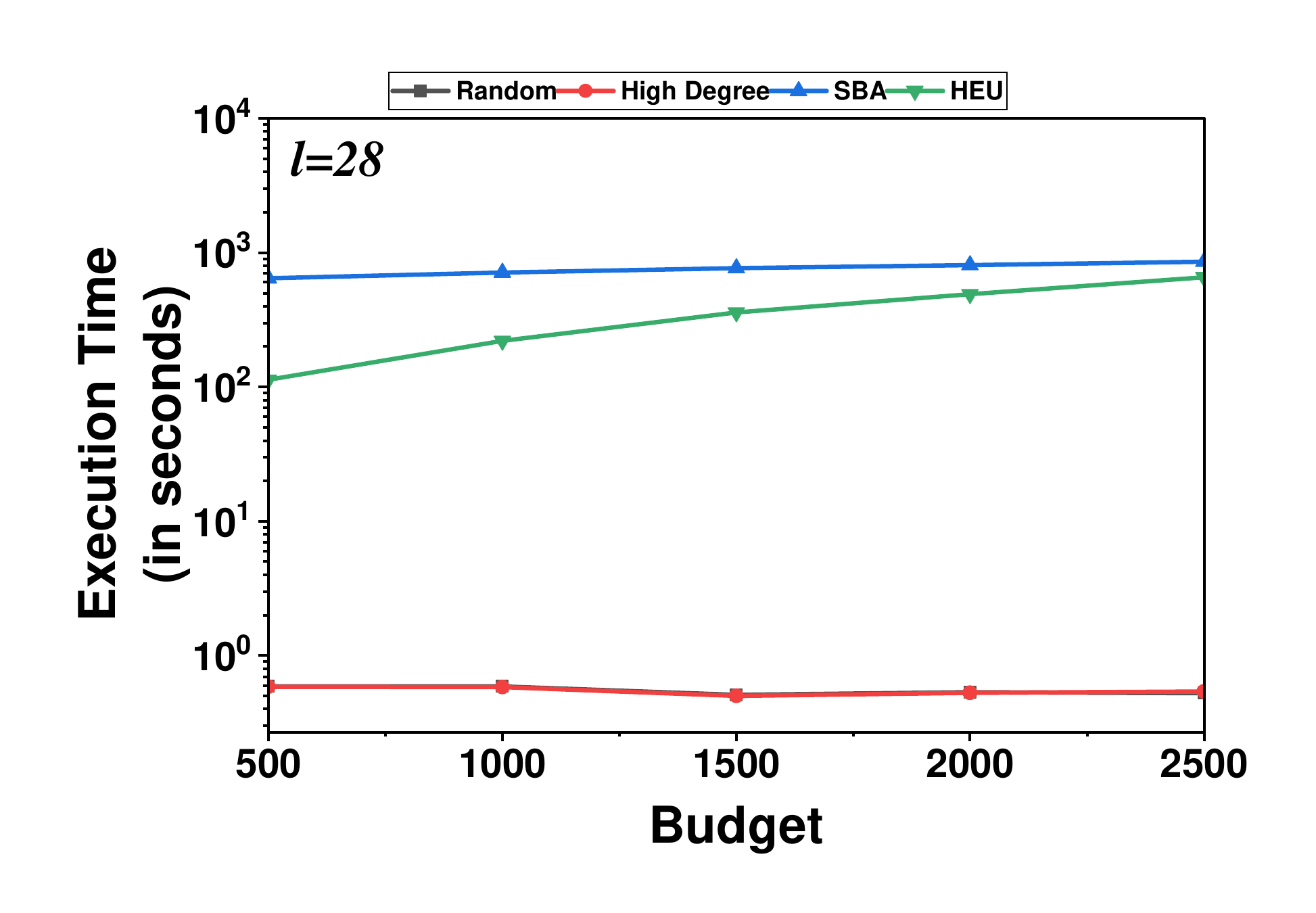} \\
(d) Trivalency
\end{tabular}
\\[6pt]

\begin{tabular}{c}
\includegraphics[width=0.2199\textwidth]{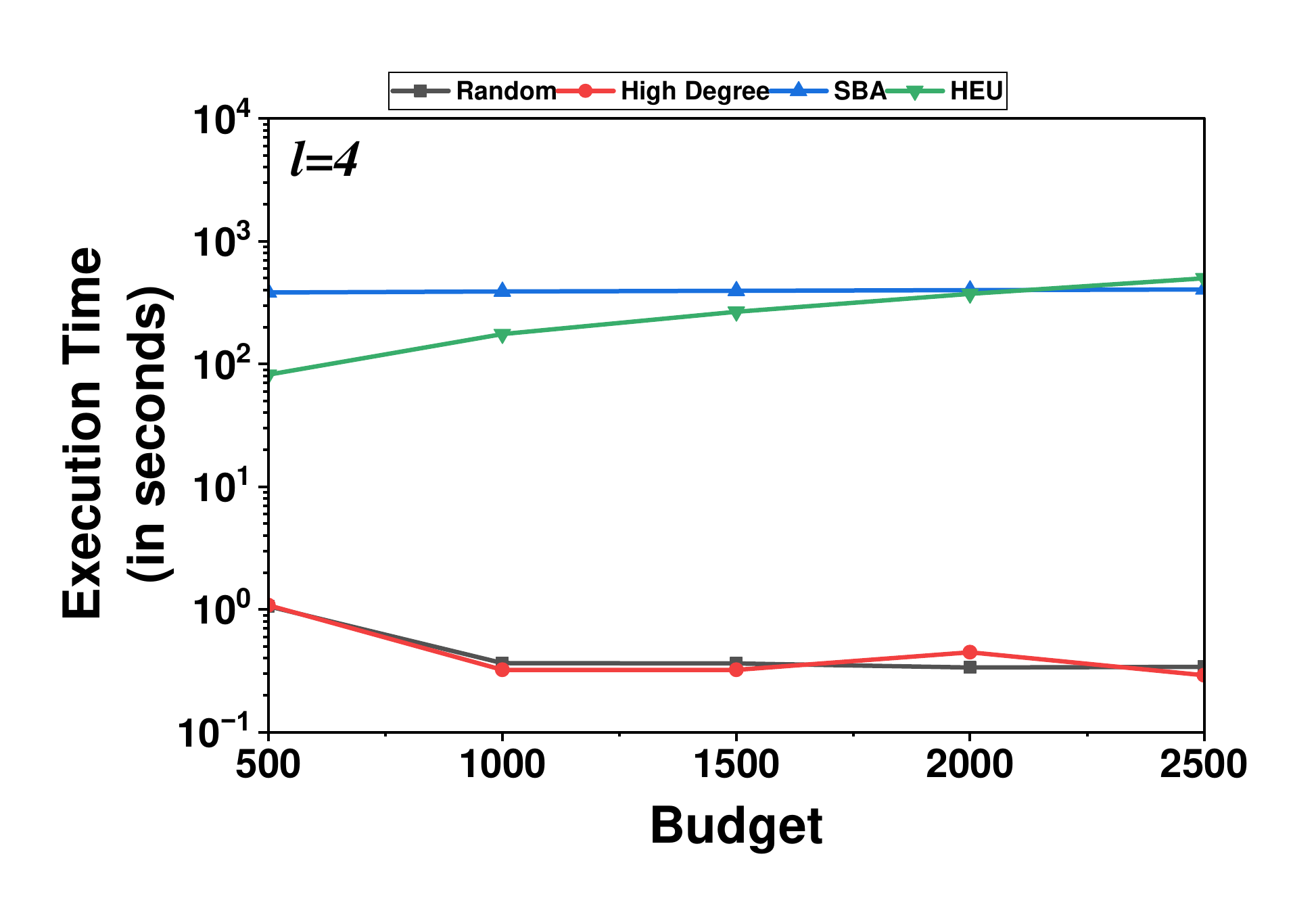} \\
(e) WC
\end{tabular} &
\begin{tabular}{c}
\includegraphics[width=0.2199\textwidth]{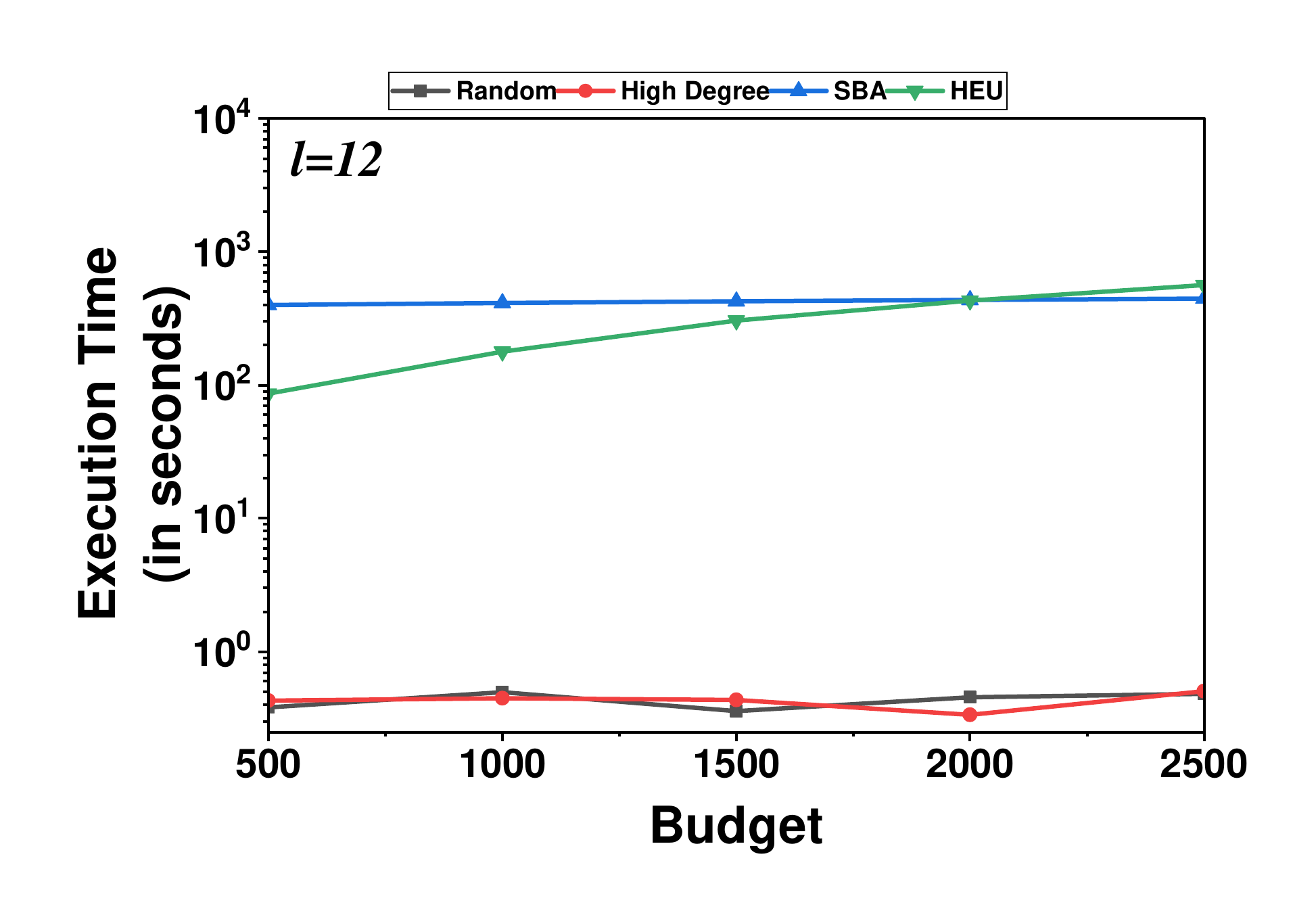} \\
(f) WC
\end{tabular} &
\begin{tabular}{c}
\includegraphics[width=0.2199\textwidth]{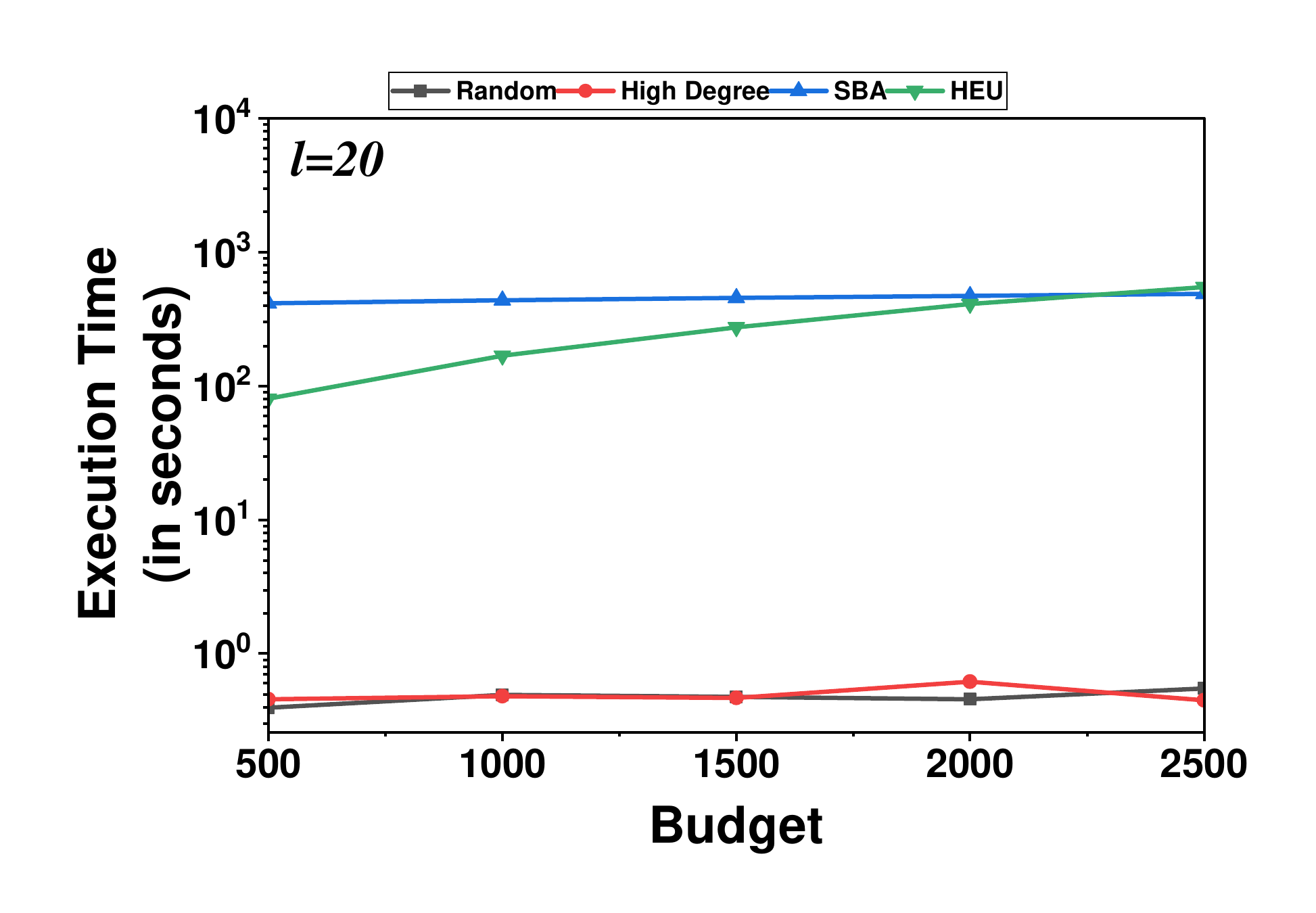} \\
(g) WC
\end{tabular} &
\begin{tabular}{c}
\includegraphics[width=0.2199\textwidth]{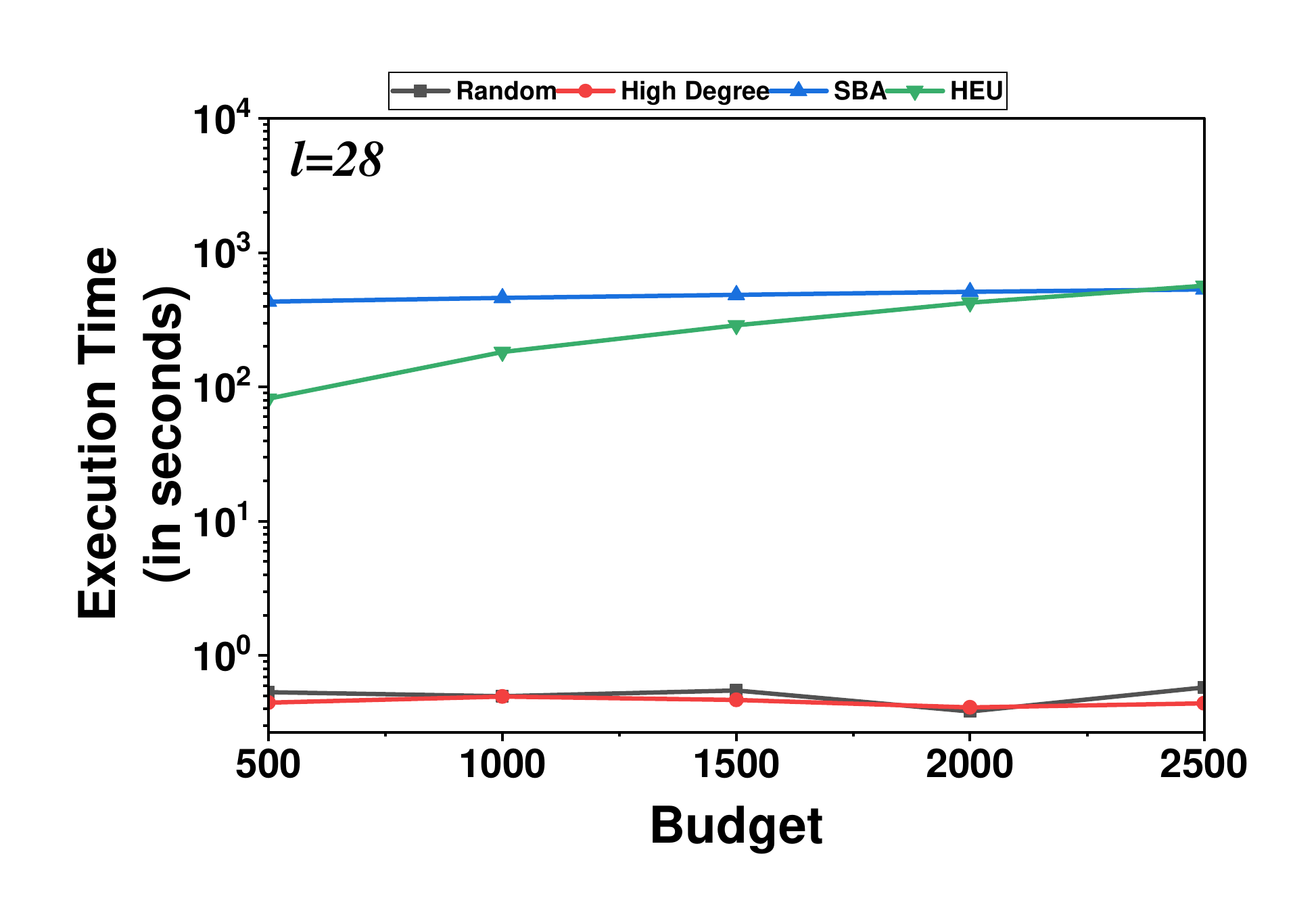} \\
(h) WC
\end{tabular}
\\[6pt]
\begin{tabular}{c}
\includegraphics[width=0.2199\textwidth]{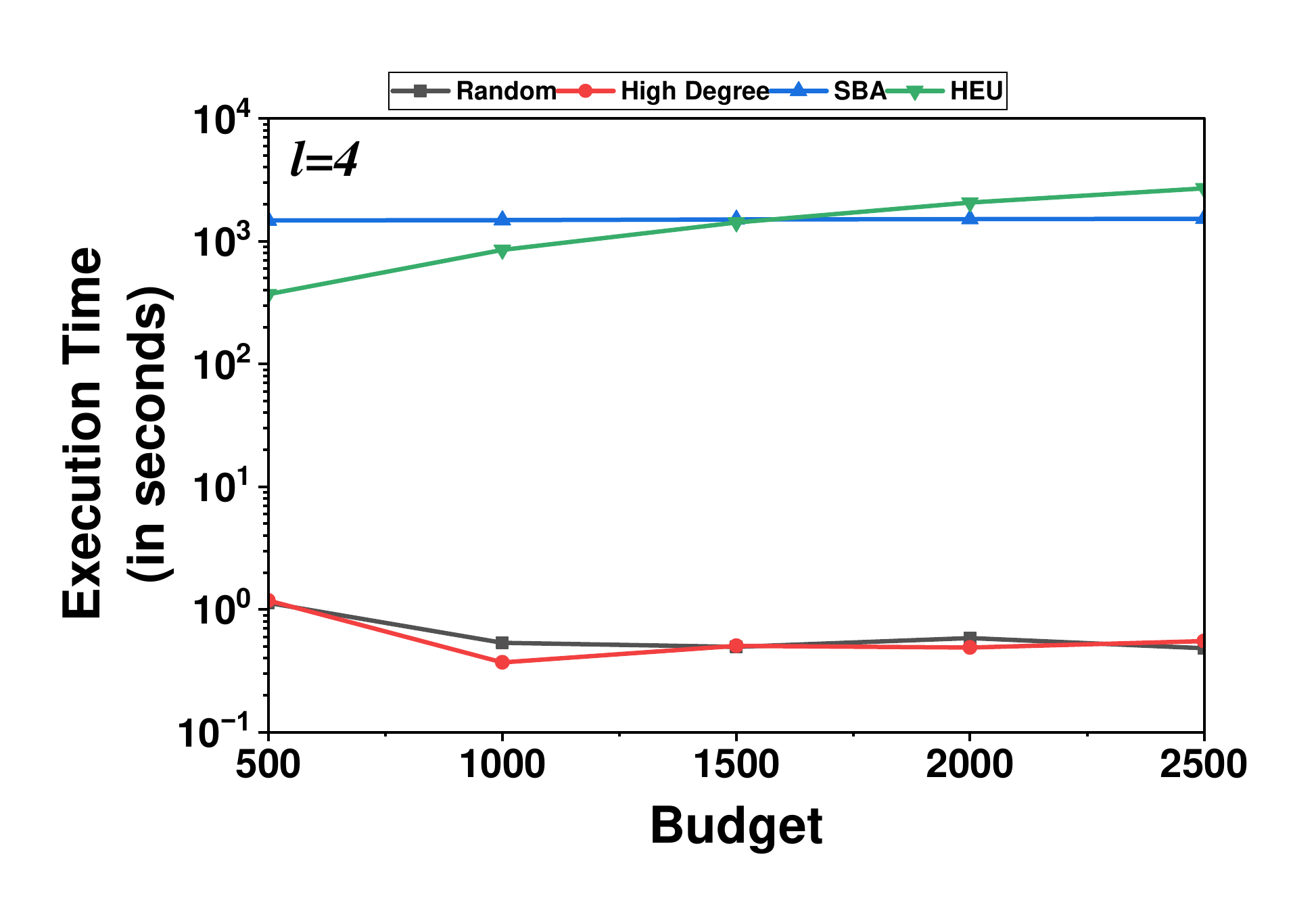} \\
(i) Trivalency
\end{tabular} &
\begin{tabular}{c}
\includegraphics[width=0.2199\textwidth]{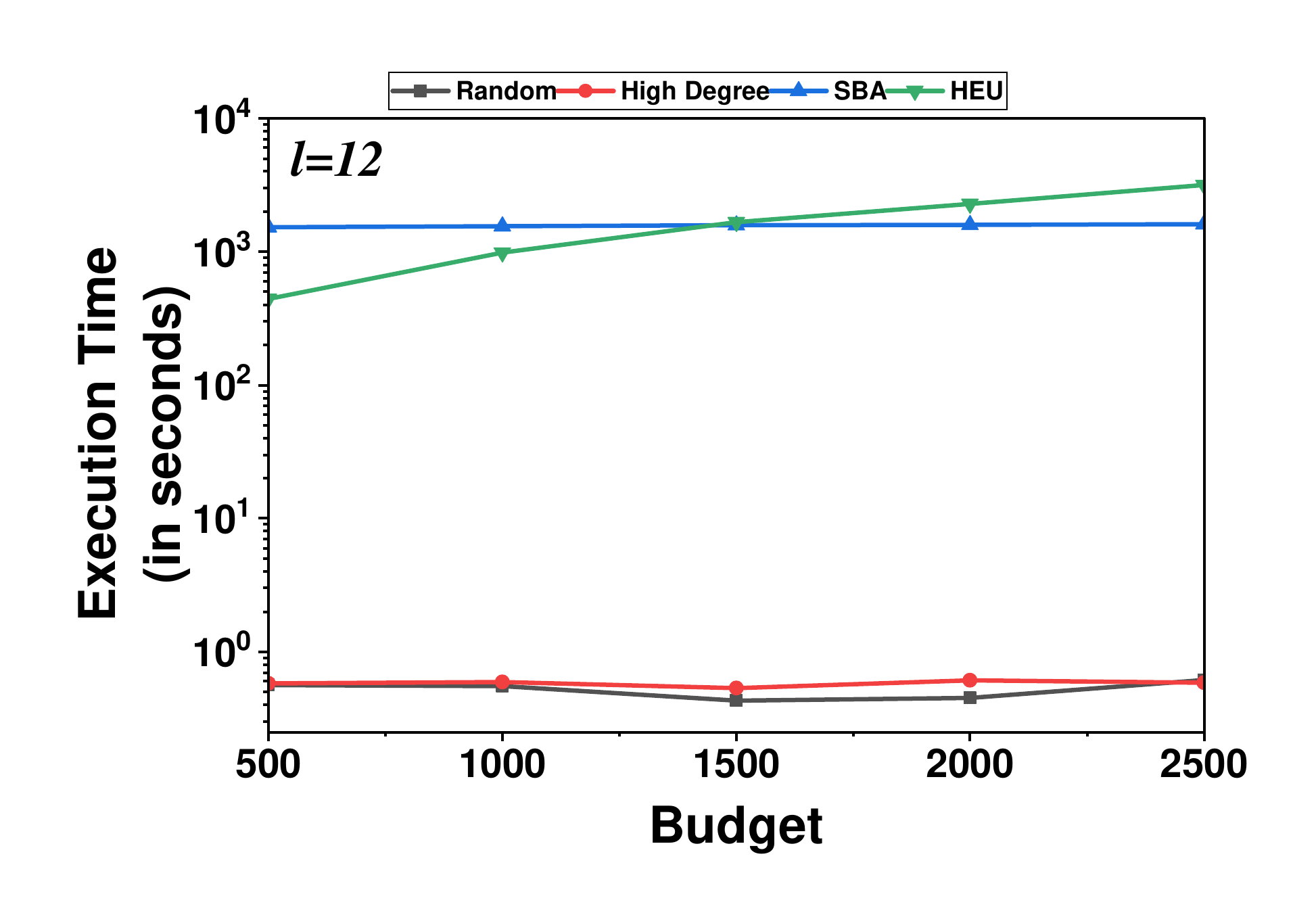} \\
(j) Trivalency
\end{tabular} &
\begin{tabular}{c}
\includegraphics[width=0.2199\textwidth]{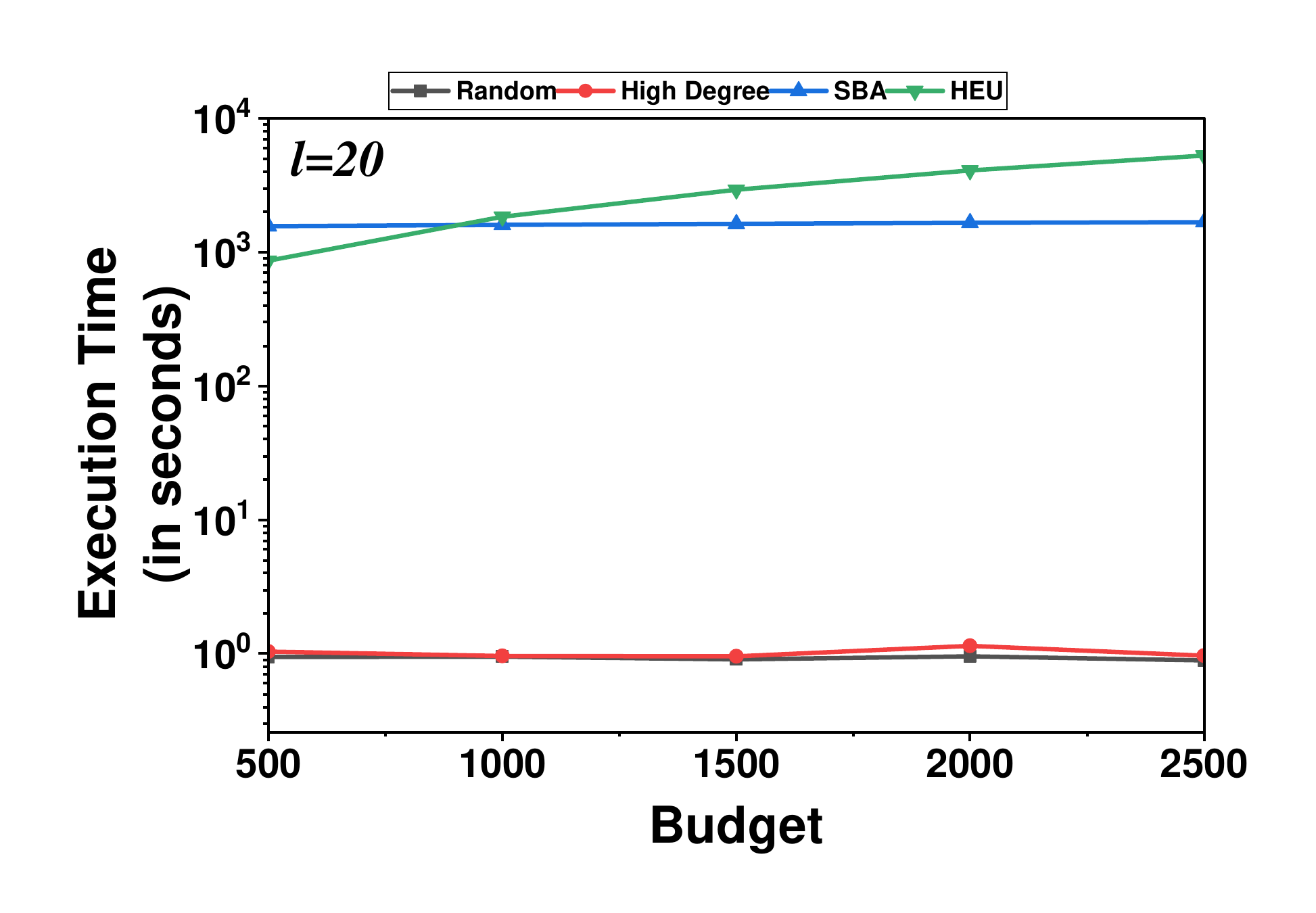} \\
(k) Trivalency
\end{tabular} &
\begin{tabular}{c}
\includegraphics[width=0.2199\textwidth]{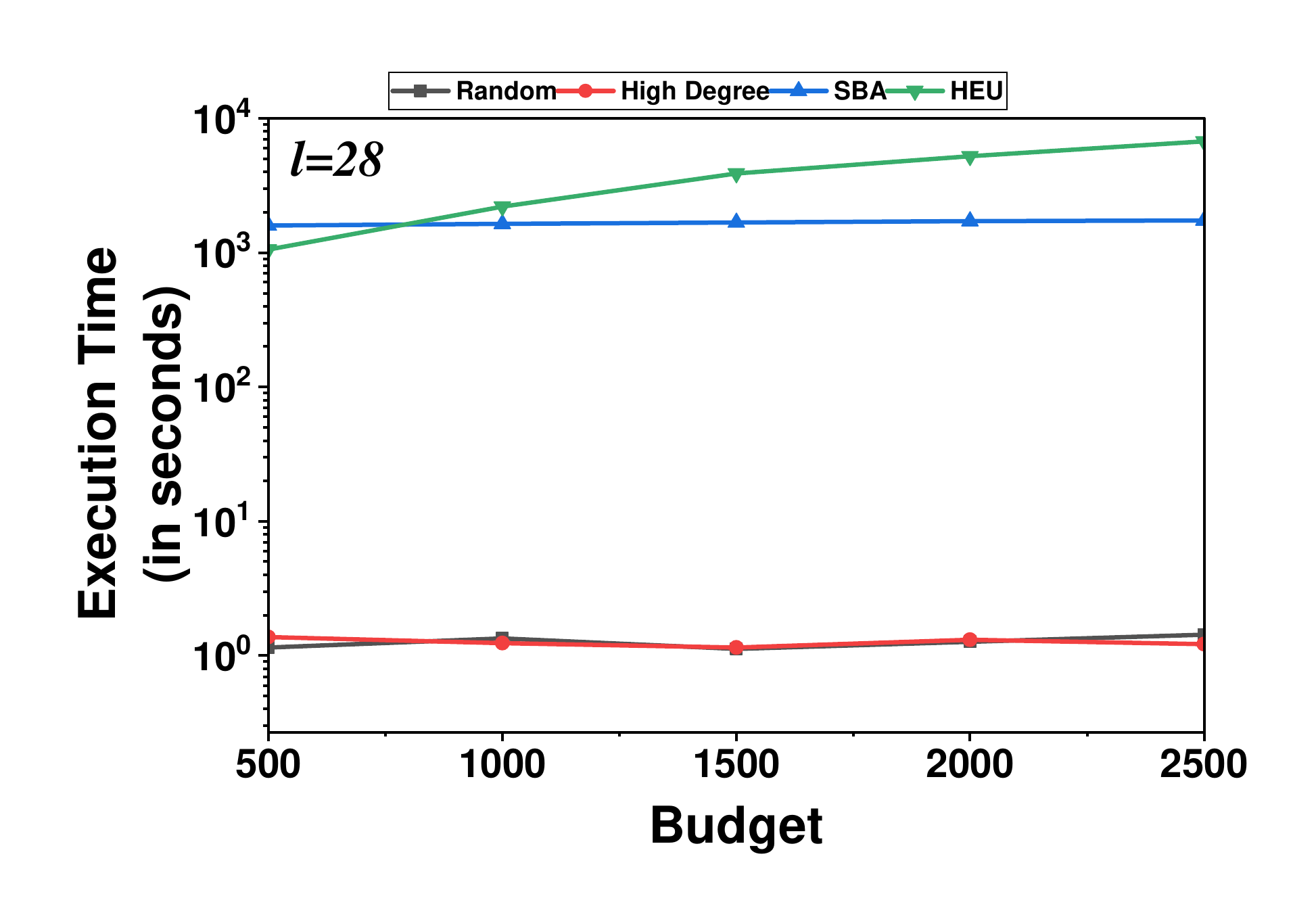} \\
(l) Trivalency
\end{tabular}
\\[6pt]

\begin{tabular}{c}
\includegraphics[width=0.2199\textwidth]{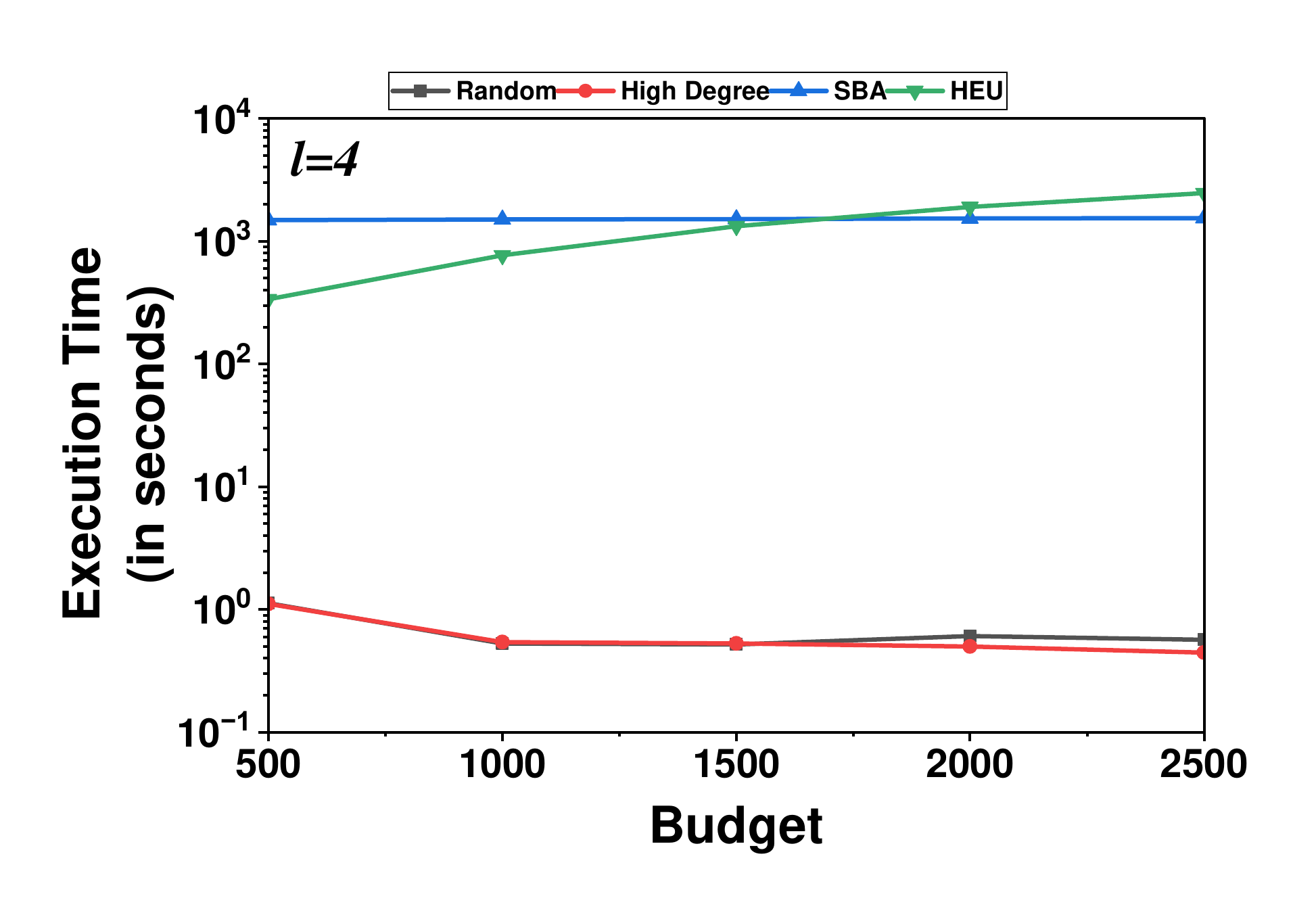} \\
(m) WC
\end{tabular} &
\begin{tabular}{c}
\includegraphics[width=0.2199\textwidth]{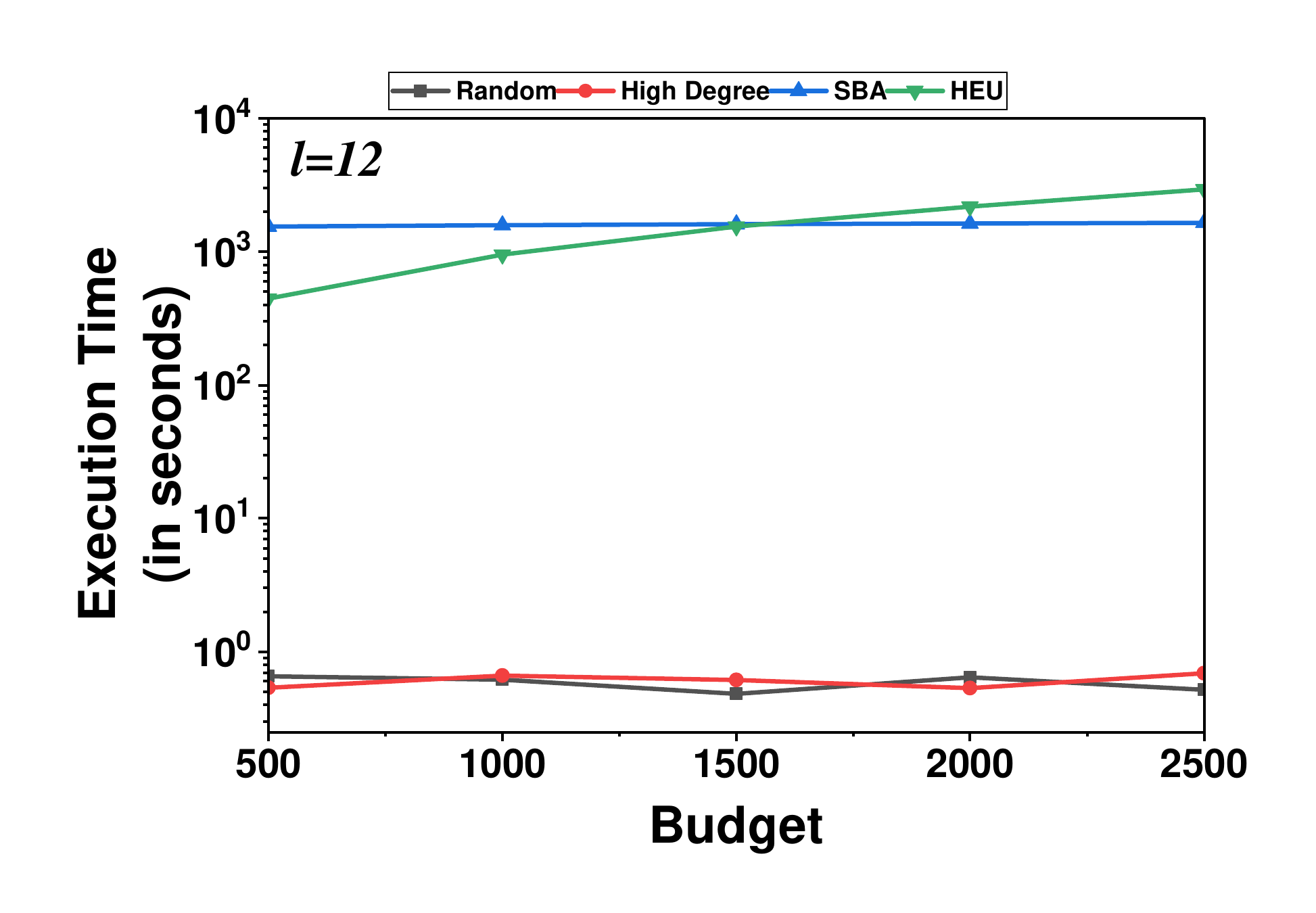} \\
(n) WC
\end{tabular} &
\begin{tabular}{c}
\includegraphics[width=0.2199\textwidth]{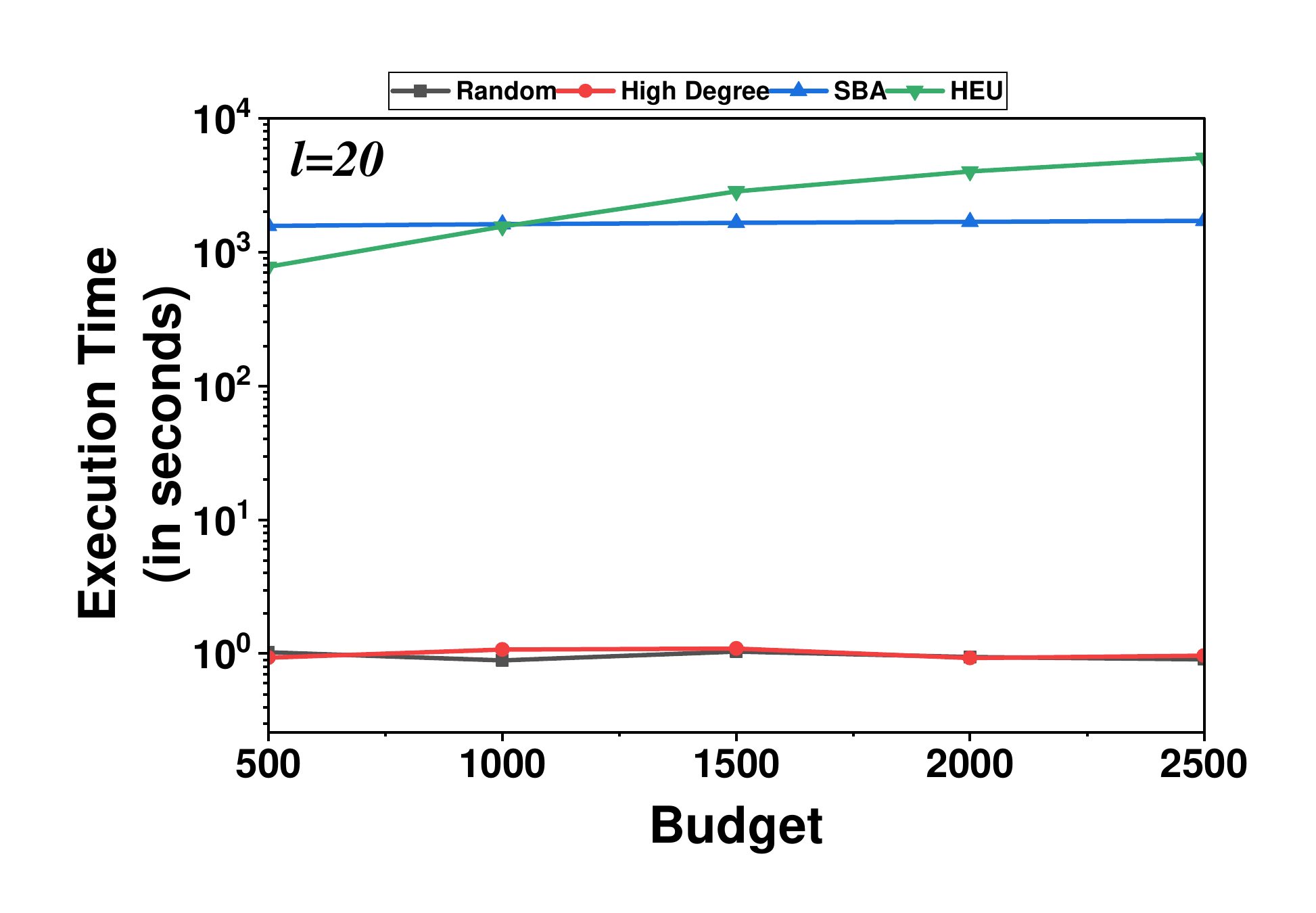} \\
(o) WC
\end{tabular} &
\begin{tabular}{c}
\includegraphics[width=0.2199\textwidth]{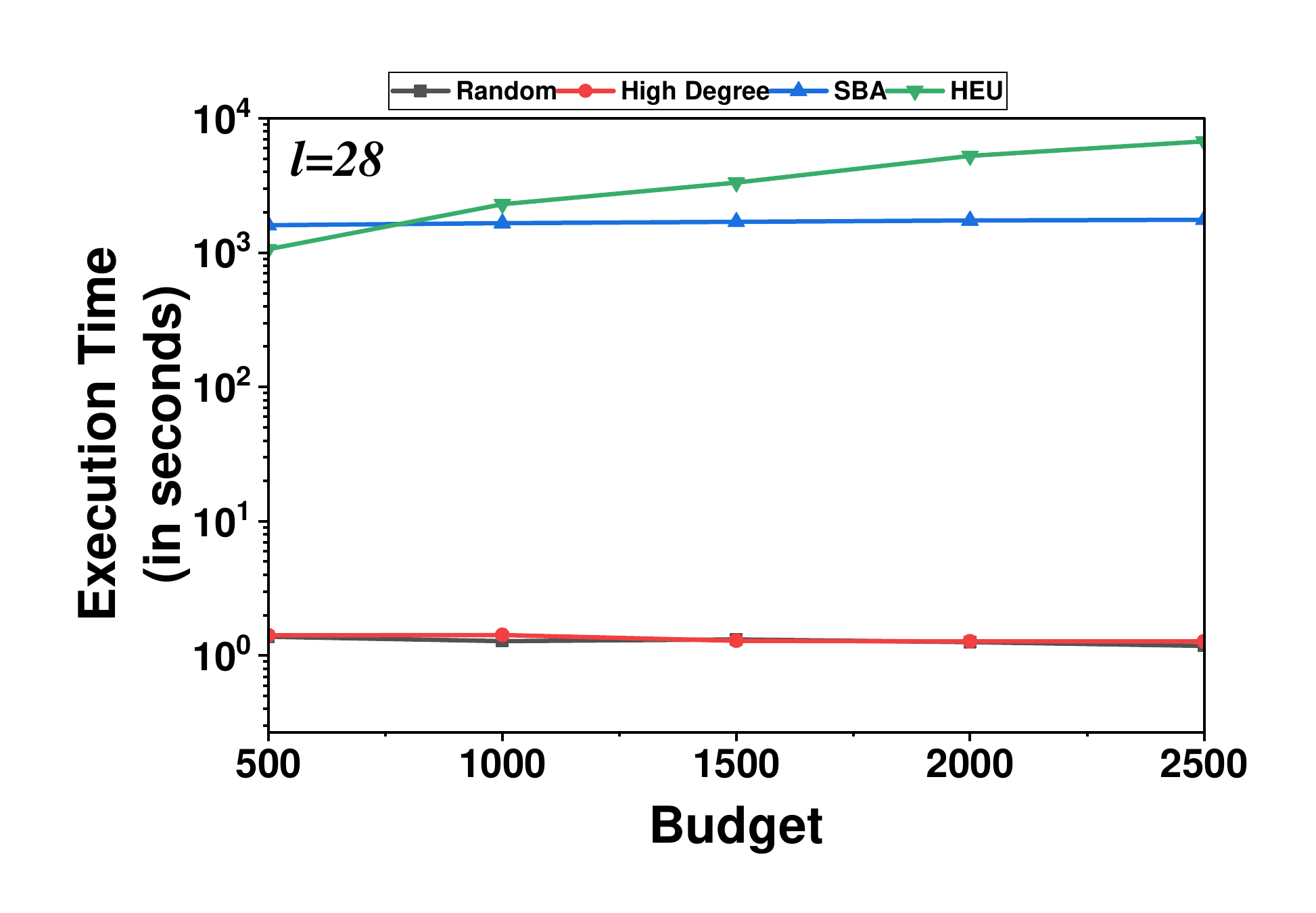} \\
(p) WC
\end{tabular}
\end{tabular}
\caption{Plots for \textit{Execution Time (in Seconds)} taken by all Algorithms under various Budgets for \textit{Euemail} ((a)-(h)) and \textit{Facebook} ((i)-(p)) Dataset}
\label{Plot2:ExecutionTime}
\end{figure}
\paragraph{\textbf{Impact on Computational Time}}
For the \textit{Euemail} dataset under trivalency settings, Random and High Degree remain fast, each running under $1$ second across all budgets and $\ell$. SBA is far slower: even at budget $500$, $\ell=4$ (Fig. \ref{Plot2:ExecutionTime}(a)) it exceeds $524$ seconds, and at budget $2500$, $\ell=28$ (Fig. \ref{Plot2:ExecutionTime}(d)) it reaches about $854$ seconds. This results from its detailed profit evaluation compared to the baselines’ simple heuristics. The same pattern appears for \textit{Euemail} under weighted cascade (Fig. \ref{Plot2:ExecutionTime}(e)–(h)). For \textit{Facebook} (trivalency) in Fig. \ref{Plot2:ExecutionTime}(i)–(l), Random and High Degree again finish under $1.5$ seconds. SBA begins near $1468$ seconds at budget $500$, $\ell=4$ and reaches almost $1737$ seconds at budget $2500$, $\ell=28$. HEU is lighter than SBA at small budgets and $\ell$, but scales steeply, exceeding $6727$ seconds at budget $2500$, $\ell=28$. Thus, Random and High Degree are fast but moderate in profit; SBA is consistently slower; and HEU—while most profitable—incurs the longest times. These trends persist for the weighted cascade case in Fig. \ref{Plot2:ExecutionTime}(m)–(p).
\section{Concluding Remarks} \label{Sec:Conclusion}
We studied the Profit Maximization in Closed Social Networks (PMCSN) Problem, in which each user can disseminate information to only a limited number of neighbors. We proposed two solutions: a sampling-based approximate algorithm and an efficient marginal-gain heuristic. For the sampling-based method, we analyzed sample and computational complexity, and for the heuristic, we provided a complexity analysis. Experiments showed how budget and $\ell$ influence profit. Future extensions will include developing learning-based approaches and targeting specific users for advertisement.

\bibliographystyle{splncs04}
\bibliography{paper}
\end{document}